\documentclass[iop, apj, onecolumn]{emulateapj-rtx4}

\usepackage{subfigure}

\shorttitle{Importance of Tides in Eccentric NS-WD Binaries}
\shortauthors{Sravan et al.}

\begin{document}

\title{Importance of Tides for Periastron Precession in \\Eccentric Neutron Star - White Dwarf Binaries}
\author{N. Sravan, F. Valsecchi, V. Kalogera}
\affil{Center for Interdisciplinary Exploration and Research in Astrophysics (CIERA), and\\
Department of Physics and Astronomy, Northwestern University, 2145 Sheridan Road, Evanston IL 60208, USA}
\and
\author{L. G. Althaus}
\affil{Grupo de Evoluci\'on Estelar y Pulsaciones, Facultad de Ciencias Astron\'omicas y Geof\'{\i}sicas, Universidad Nacional de La Plata, Paseo
del Bosque s/n, (1900) La Plata, Argentina
Instituto de Astrof\'{\i}sica La Plata, CONICET-UNLP, Argentina}

\begin{abstract}

Although not nearly as numerous as binaries with two white dwarfs, eccentric neutron star-white dwarf (NS-WD) binaries are important gravitational-wave (GW) sources for the next generation of space-based detectors sensitive to low frequency waves. Here we investigate periastron precession in these sources as a result of general relativistic, tidal, and rotational effects; such precession is expected to be detectable for at least some of the detected binaries of this type. Currently, two eccentric NS-WD binaries are known in the galactic field, PSR J1141-6545 and PSR B2303+46, both of which have orbits too wide to be relevant in their current state to GW observations. However, population synthesis studies predict the existence of a significant Galactic population of such systems. {Though small in most of these systems, we find that tidally induced periastron precession becomes important when tides contribute to more than 3\% of the total precession rate. For these systems, accounting for tides when analyzing periastron precession rate measurements can improve estimates of the WD component mass inferred and, in some cases, will prevent us from misclassifying the object. However, such systems are rare due to rapid orbital decay. To aid the inclusion of tidal effects when using periastron precession as a mass measurement tool, we derive a function that relates the WD radius and periastron precession constant to the WD mass.}

\end{abstract}

\section{Introduction} \label{Intro}

Future, space-based {GW} observations will target Galactic binaries in tight orbits as primary sources of continuous emission (e.g. LISA, Danzmann \& the LISA study team 1996, Hughes 2006, and eLISA/NGO, Amaro-Seoane et al. 2012). Among those expected to be detected at relatively high signal-to-noise ratio are some eccentric binaries, whose periastron precession will leave an imprint on the GW signal \citep{wvk08}. In this paper our focus lies on the population of eccentric neutron star - white dwarf binaries (NS-WD); two such systems are currently observed as radio pulsars orbiting WDs.

{Periastron precession in eccentric binaries is caused by a combination of tidal, rotational and general relativistic (GR) effects. Tides and rotation produce distortions in the binary components which perturb the stellar gravitational potential from its pure Newtonian form and the orbit from its Keplerian form, driving precession.} Both these contributions depend on the internal mass distributions (through the, so-called, periastron precession constant $k_2$), masses, and radii of the components, along with the orbital period and eccentricity. The GR contribution, on the other hand, depends solely on the total mass of the system and its orbital elements. Periastron precession is detectable through its imprint on the GWs emitted by these sources. In the absence of periastron precession, the GW radiation is emitted at multiples $n$ of the orbital frequency $\nu_{\rm orb}$. Periastron precession causes each of these harmonics to split into triplets with frequencies $n\nu_{\rm orb} \pm \dot{\gamma}/\pi$ and $n\nu_{\rm orb}$, where $\dot{\gamma}$ is the periastron precession rate (e.g., Willems et al. 2008).

\citet{s01} suggests that, if the orbital elements are known, one can use $\dot{\gamma}$ to extract the total mass of low frequency eccentric binaries assuming that GR effects dominate the periastron precession rate. However, \citet{wvk08} use polytropic models to investigate periastron precession in eccentric double WD (DWD) binaries and demonstrate that the tidal and rotational distortions of the WD components can significantly affect such a precession in short-period binaries. This implies that, ignoring the tidal and rotational contributions, when interpreting periastron precession measurements from these binaries, could lead to an overestimate of the total system mass extracted. Furthermore, they anticipate that accounting for all three contributions would entail degeneracies, given the dependency of $\dot{\gamma}$ on the internal structure ($k_2$) and radius of both components. \citet{vfw11} use detailed WD models to study periastron precession in eccentric DWDs and demonstrate that the components' masses could be overestimated by order of magnitudes if tides are not properly taken into account. They also show that there exists a correlation between $k_2R^5$ and the WD mass that allows the use of the periastron precession rate to place constraints on some combination of the components' masses\footnote{If GR is the dominant mechanism driving periastron precession, the total system mass can be determined. Instead, if tides are the dominant mechanism, constraints can be placed on $(M_1+M_2)^{-5/3}[k_1R_1^5(M_2/M_1)+k_2R_2^5(M_1/M_2)]$, where the term $k_iR_i^5$ for each WD component ($1, 2$) is a function of its mass according to Eq. (7) in \citet{vfw11}.} at any orbital frequency.

Unlike DWDs, where both components contribute to periastron precession, NS-WD binaries are a much cleaner probe of WD physics. In such systems, the periastron precession rate carries the sole signature of the WD because the tidal and rotational distortions of the NS contribute negligibly. Furthermore, the theoretically predicted formation rate of galactic eccentric NS-WD binaries is between 10 to 10,000 times the expected formation rates of eccentric DWDs (e.g., Kalogera et al. 2005 and references therein, Willems et al. 2007). Additionally, since pulsar-timing measurements could yield an independent measurement of the components' masses, these systems may be used to test our predictions and models from analyzing periastron precession rates.

Two eccentric NS-WD binaries, PSR J1141-6545 \citep{klm00} and PSR B2303+46 \citep{kk99}, have been discovered in the Milky Way. These binaries have lent support to the existence of a non-traditional formation mechanism for NS-WD binaries (Portgies Zwart \& Yungelson 1999; Tauris \& Sennels 2000; Nelemans, Yungelson, \& Portegies Zwart 2001; Brown et al. 2001; Davies, Ritter, \& King 2002). The traditional mechanism leads to circular binaries: the more massive primary evolves faster to become a NS, followed by a mass transfer phase from the secondary (WD progenitor) that circularizes any eccentricity introduced by the supernova mechanism that led to the formation of the NS. However, if the progenitors are both massive enough to evolve into massive WDs, the following scenario may unfold. The primary star evolves into a WD after a phase of mass transfer to the secondary. If the secondary acquires enough mass to evolve into a NS, a common envelope forms. During the common envelope phase, the NS progenitor loses its envelope, leaving behind its naked He core in a tight orbit with the WD companion. Subsequently, asymmetries in the NS formation process kick the resulting NS-WD into an eccentric orbit. Apart from this, several other evolutionary mechanisms involving multiple mass transfer and common envelope phases have also been proposed (\citet{cbt06}). Nonetheless, all such mechanisms require the NS-forming supernova explosion as the final step to impart eccentricity to the orbit.

Following \citet{vfw11}, our goal is to investigate the importance of WD tides in driving periastron precession in eccentric NS-WD binaries, in order to facilitate accurate mass interpretation from periastron precession rate measurements. In what follows we also take into account the contribution to periastron precession due to rotation. However, similarly to \citet{vfw11}, we find that periastron precession is primarily tidally-induced at frequencies were GR is no longer important. 

The plan of the paper is as follows. In \S~\ref{theory} we outline the equations governing the tidal, rotational and GR contributions to periastron precession. In \S~\ref{observed} we analyze the importance of tides in driving periastron precession in the observed eccentric NS-WD binary systems. In \S~\ref{popsynth} we investigate periastron precession in the population of eccentric NS-WD binaries in the Milky Way predicted by population synthesis studies to understand the role of tides. We conclude in \S~\ref{conclusions}.

\section{Physical Processes Driving Periastron Precession} \label{theory}

We consider an eccentric NS-WD binary system containing a NS of mass $M_{_{NS}}$, and a WD of mass $M_{_{WD}}$, radius $R_{_{WD}}$, uniformly rotating with angular velocity $\Omega_{_{WD}}$. We take the NS to be a point mass. We assume that the axis of rotation of the WD is perpendicular to the orbital plane. Let $P$ be the period of the orbit, $a$ the semi-major axis, and $e$ the orbital eccentricity. {For simplicity, we take tides to be quasi-static (the regime where the orbital and rotational periods are long compared to the free oscillation modes of the stars, Cowling 1938; Sterne 1939; Smeyers \& Willems 2001), but we note that investigations targeting periastron precession in non-degenerate stars show that the effects of dynamic tides become more significant as the orbital and/or rotational period, and eccentricity increase (\citet{sw01}; \citet{wc02}; \citet{wc05}). We also note that while \citet{wdk10} demonstrate that GR dominates over quasi-static tides in driving the evolution of the orbital separation and eccentricity in NS-WD binaries, recent investigations targeting dynamic tides in binaries hosting a WD and another compact object find that they can significantly speed up the orbital and spin evolution (\citet{bqa13}; \citet{fl11})}. The contribution to the periastron precession rate, $\dot{\gamma}$, due to quadrupole tides raised in the WD is \citep{s39}

\begin{equation} \label{tid}
\dot{\gamma}_{_{Tid,WD}}= \frac{30\pi}{P} \left(\frac{R_{_{WD}}}{a}\right)^5 \frac{ M_{_{NS}}}{M_{_{WD}}} \frac{1+\frac{3}{2} e^2+\frac{1}{8}e^4} {(1-e^2)^5} k_2.
\end{equation}

{Here, $k_2$, also known as the quadrupolar periastron precession constant, is a measure of the WD's central concentration and is given by}

\begin{equation} \label{k2}
{2k_2=\frac{\xi^*_{_{WD,T}}\left(R_{_{WD}}\right)}{R_{_{WD}}}-1}
\end{equation}

{Here $\xi^*_{_{WD,T}} = \xi_{_{WD,T}}/(\epsilon_{_{T}} c_{200}))$, where $\xi_{_{WD,T}}$ denotes the radial component of the tidal displacement of a mass element of the WD and $c_{200}$ is the Fourier coefficient associated with the l=2, m=0, and k=0 term in the spherical harmonic expansion of the tide-generating potential \citep{ps90}}. The values of $k_2$ range from 0 for a point mass to 0.74 for an equilibrium sphere with uniform density. In the quasi-static tides regime, $\xi_{_{WD,T}}$ is a solution to the homogeneous second-order differential equation (e.g. Smeyers \& Willems 2001)

\begin{equation} \label{de}
\frac{d^2\xi_{_{WD,T}}(r)}{dr^2}+2\left(\frac{1}{g(r)}\frac{dg(r)}{dr}+\frac{1}{r}\right)\frac{d\xi_{_{WD,T}}(r)}{dr}-\frac{l(l+1)-2}{r^2}\xi_{_{WD,T}}(r)=0
\end{equation}

where $g$ denotes the local gravity. The solution to Eq. (\ref{de}) must remain finite at $r=0$ and $r=R_{_{WD}}$, and must satisfy the following boundary condition at the WD surface

\begin{equation} \label{bc}
\left(\frac{d\xi_{_{WD,T}}(r)}{dr}\right)_{R_{_{WD}}}+\frac{l-1}{R_{_{WD}}}\xi_{_{WD,T}}(R_{_{WD}})=\epsilon_{_{T}}(2l+1)c_{_{l,0,0}}.
\end{equation}

Here, $\epsilon_{_{T}}=(R_{_{WD}}/a)^3(M_{_{NS}}/M_{_{WD}})$ indicates the strength of the tidal force versus gravity at the WD's equator, $l$ is the longitudinal mode in the spherical harmonic expansion of the tide-generating potential, and $c_{_{l,0,0}}$ are Fourier coefficients of degree $l$. Since $c_{_{l,m,k}}$ depend on the WD radius and the semi-major axis as $(R_{_{WD}}/a)^{l-2}$, investigations of quasi-static tides in these sources are often restricted to the dominant $l=2$ terms. Here we only consider $l=2$.

Rotation contributes to the periastron precession rate through the quadrupole distortion of the gravitational field caused by the centrifugal force \citep{s39} and it is given by
\begin{equation} \label{rot}
\dot{\gamma}_{_{Rot,WD}}= \frac{2\pi}{P} \left(\frac{R_{_{WD}}}{a}\right)^5 \frac{M_{_{NS}}+M_{_{WD}}}{M_{_{WD}}} \frac{(\Omega_{_{WD}}/\Omega)^2} {(1-e^2)^2} k_2,
\end{equation}
where $\Omega = 2\pi/P$ is the mean motion. In this work, we assume synchronization at periastron for the spin of the WD component. This assumption leads to tides being dominant over rotation in driving periastron precession at any orbital frequency. The contribution due to rotation further decreases for a subsynchronous WD component. As the current observational constraints on rotation rates in single and double WD suggest that WDs are slow rotators with synchronization at periastron being an upper limit on WD rotation rate (see the discussion in \S~3.2 of \citet{vfw11}, and references therein), we can safely assume that periastron precession is primarily tidally-induced at frequencies where GR is no longer important. 

Finally, the GR contribution to the periastron precession rate to the leading quadrupole order is given by
\begin{equation} \label{gr}
\dot{\gamma}_{_{GR}}= \left(\frac{30\pi}{P}\right)^{5/3} \frac{3 G}{c^2}\frac{M_{_{NS}}+M_{_{WD}}} {a(1-e^2)}
\end{equation}
where $G$ is the gravitational constant and $c$ is the speed of light \citep{lc37}.
The total periastron precession rate is the sum of the tidal, rotational and GR contributions.

Apart for causing periastron precession, tides and GR also induce orbital evolution. {As we take tides to be quasi-static, here we only consider the evolution of orbital elements due to GR (\citet{wdk10}; but see also \citet{bqa13}; \citet{fl11})}. The time average of the rate of change of $a$ and $e$ due to GR are given by \citep{p64}
\begin{equation} \label{dadt}
\left<\frac{da}{dt}\right>= - \frac{64}{5}\frac{G^3M_{_{NS}}M_{_{WD}}(M_{_{NS}}+M_{_{WD}})}{c^5a^3(1-e^2)^{7/2}}\left(1+\frac{73}{24}e^2+\frac{37}{96}e^4\right)
\end{equation}
\begin{equation} \label{dedt}
\left<\frac{de}{dt}\right>= - \frac{304}{15}e\frac{G^3M_{_{NS}}M_{_{NS}}(M_{_{NS}}+M_{_{NS}})}{c^5a^4(1-e^2)^{5/2}}\left(1+\frac{121}{304}e^2\right).
\end{equation}
From Eqs. (\ref{dadt}) and (\ref{dedt}) it can be derived that
\begin{equation} \label{dade}
\left<\frac{da}{de}\right>= \frac{12}{19}\frac{a}{e}\frac{[1+(73/24)e^2+(37/96)e^4]}{(1-e^2)[1+(121/304)e^2]}.
\end{equation}

We will now make use of the above formulation to study periastron precession in eccentric NS-WD binaries. In what follows, we refer to the radius of the WD as $R$.

\begin{deluxetable}{lrrrrrrr}
\tabletypesize{\scriptsize}
\tablecaption{Properties of The Two Observed Eccentric NS-WD Binaries \label{sys_params}}
\tablewidth{0pt}
\tablehead{\colhead{Name} & \colhead{$M_{_{NS}}(M_{\odot})$} & \colhead{$M_{_{WD}}(M_{\odot})$} & \colhead{$a(R_{_\sun})$} & \colhead{$P (hr)$} & \colhead{$P_{spin}(ms)$} & \colhead{$e$} & \colhead{$t_{cool}(Myr)$}}
\startdata
PSR J1141-6545 & 1.27\tablenotemark{1} & 1.02\tablenotemark{1} & 1.89\tablenotemark{2} & 4.744\tablenotemark{3} &  393.9\tablenotemark{3,4} & 0.17\tablenotemark{1} & 1.45\tablenotemark{3,4} \\
PSR B2303+46 & 1.40\tablenotemark{3} & 1.24\tablenotemark{2} & 31\tablenotemark{2} & 296.2\tablenotemark{3,4} & 1066\tablenotemark{3,4} & 0.658\tablenotemark{3} & 29.7\tablenotemark{3,4} \\
\enddata
\tablenotetext{1}{\citet{bbv08}}
\tablenotetext{2}{\citet{drk02}}
\tablenotetext{3}{\citet{kkl04}}
\tablenotetext{4}{\citet{kk99}}
\end{deluxetable}

\section{The Observed Eccentric NS-WD Binaries: PSR J1141-6545 and PSR B2303+46} \label{observed}

There are currently two known eccentric binary radio pulsars with WD companions that are believed to have formed via the mechanism described in \S~\ref{Intro}: PSR J1141-6545 and PSR B2303+46. A summary of their properties relevant to this analysis is in Table \ref{sys_params}. At present, neither of these systems are verification binaries for either LISA or eLISA/NGO. However, as these systems spiral-in due to GW emission, their orbital frequencies increase. Here we examine whether PSR J1141-6545 and PSR B2303+46 still preserve an eccentricity when they evolve into the LISA sensitivity band (the sensitivity planned for LISA is from $10^{-4}$ Hz to $0.1$ Hz: henceforth referred to as ``The LISA band'', and that for eLISA/NGO is from $10^{-4}$ Hz to $1$ Hz). Moreover, if they do, we investigate the importance of tidally- and rotationally-induced periastron precession as they evolve throughout it.

\begin{figure}
\epsscale{1.17}
\plottwo{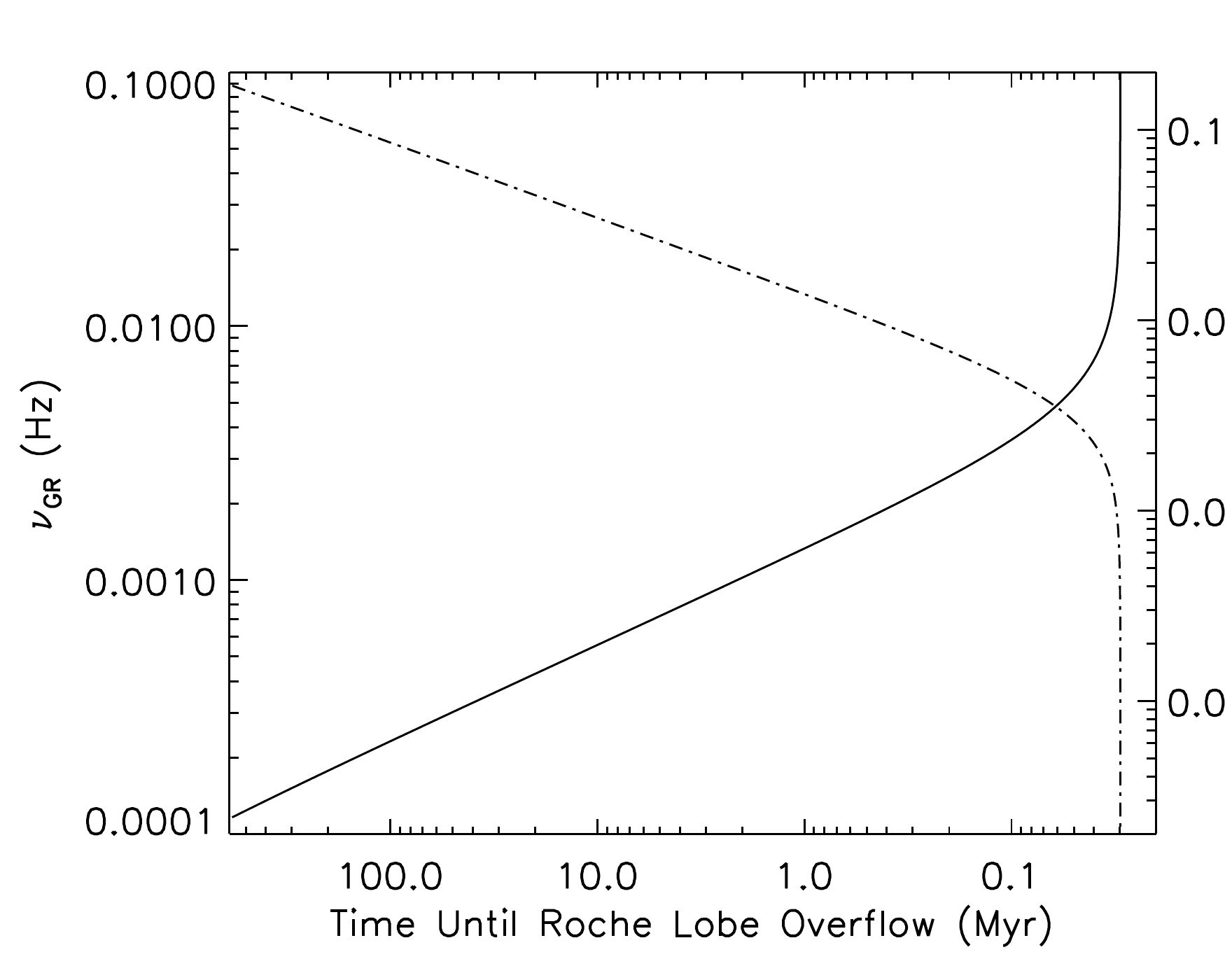}{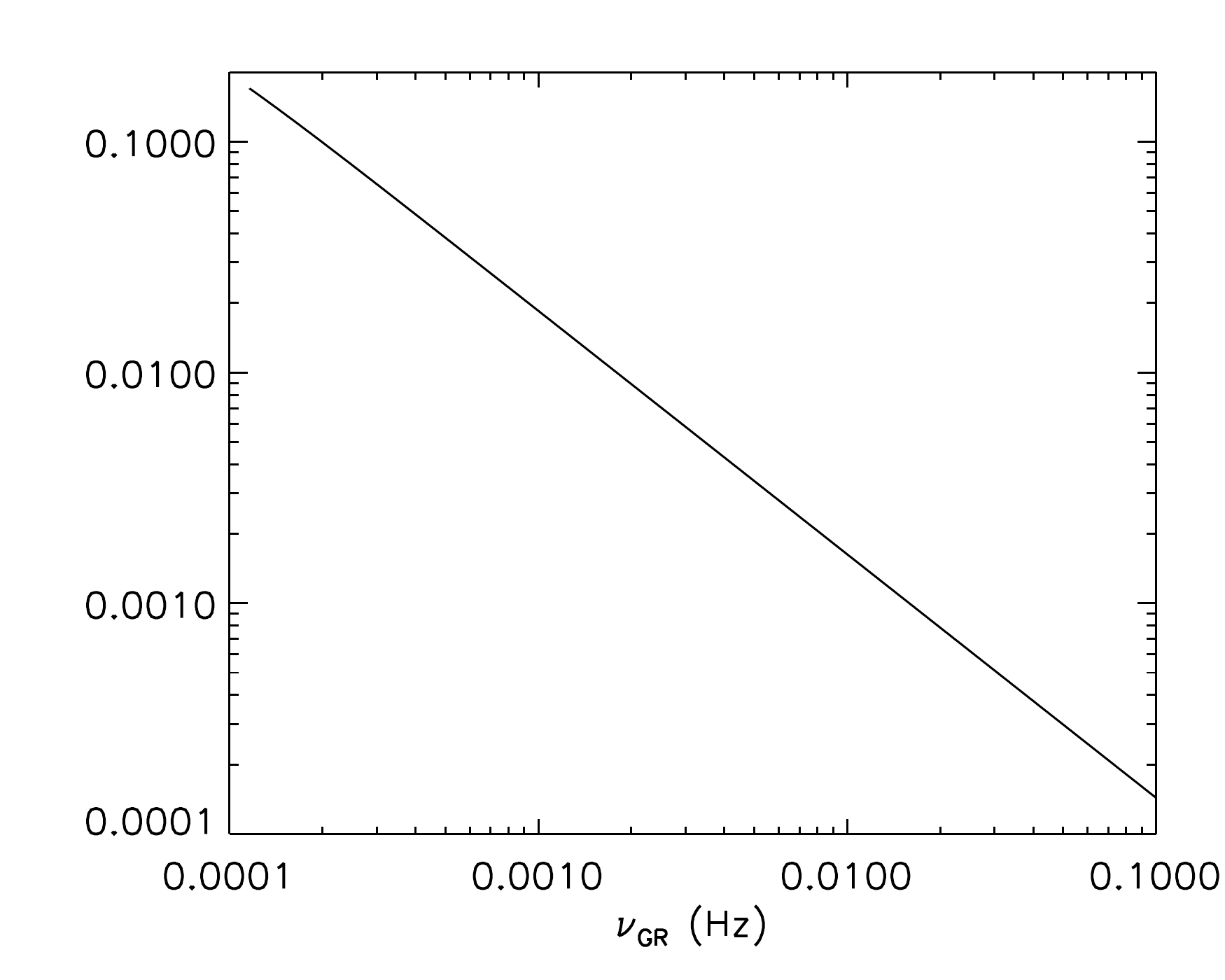}
\caption{{{\it Left}: Evolution of the GW frequency ($\nu_{GR}$, solid line) and eccentricity ($e$, dot-dashed line) of PSR J1141-6545 as a function of time until Roche lobe overflow. {\it Right}: Evolution of eccentricity as a function of GW frequency of PSR J1141-6545.}}
\label{J1141enutime}
\end{figure}

We consider first PSR J1141-6545. We compute the evolution of {its} GW frequency ($\nu_{GR} = 2\nu_{orb}$) and eccentricity as a function of time {until Roche lobe overflow} using Eqs. (\ref{dadt}) and (\ref{dedt}). The results shown in Fig. \ref{J1141enutime} demonstrate that this system will still exhibit a small but non-zero eccentricity as it evolves throughout the LISA band. Next, we simultaneously evolve the stellar and orbital parameters to analyze the contribution of tides, rotation and GR to periastron precession, as it evolves throughout the LISA band. To model the WD component we use 1.06 $M_{\sun}$ O$\backslash$Ne$\backslash$Mg (ONeMg) WD models (Althaus et al. 2007: these models are described later in detail in Section \S~\ref{properties}). At this stage we use Eqs. (\ref{dadt}) and (\ref{dedt}) and calculate $k_{2}R^{5}$ as a function of the WD age for a sequence of WD models using Eqs. (\ref{k2}), (\ref{de}) and (\ref{bc}). We then interpolate between the computed values.

\begin{figure}
\epsscale{0.65}
\plotone{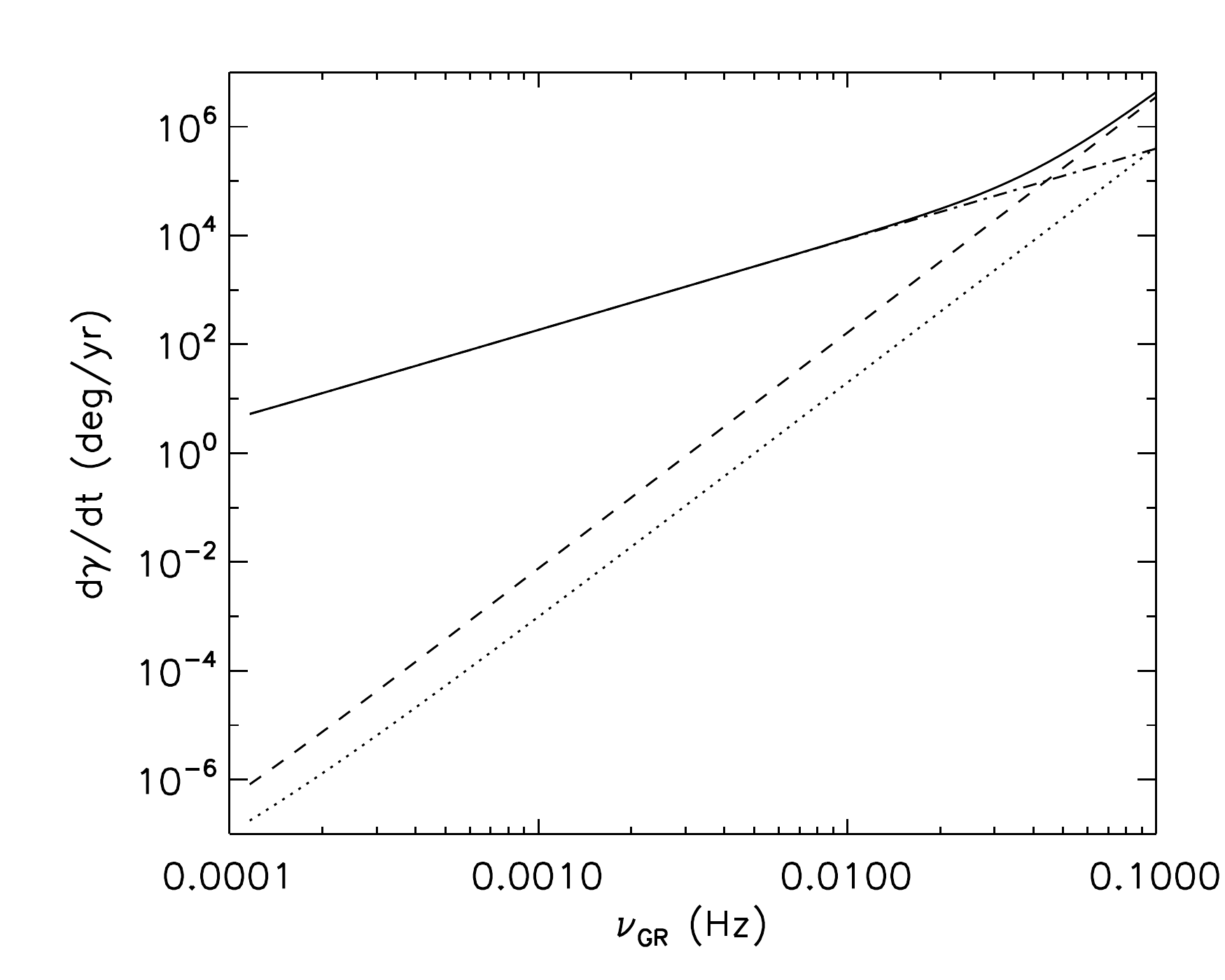}
\caption{{Periastron precession rates as a function of GW frequency for PSR J1141-6545, assuming that the WD component is old at present. The dashed line indicates the tidal contribution, the dotted line indicates the rotational contribution and the dot-dashed line indicates the GR contribution. The solid line represents the total periastron precession rate. In a binary like PSR J1141-6545, the WD component will overfill its Roche lobe at $\nu_{GR} = 0.114$ Hz.}}
\label{J1141}
\end{figure}

{The results of the time evolution of PSR J1141-6545 are shown in Fig. \ref{J1141}. Here we show the individual contributions to periastron precession as a function of the GW frequency assuming that the WD component is old (7.6 Gyr) at present. Our results are not significantly affected by the WD evolutionary stage (see \S~\ref{properties}). The trends in Fig. \ref{J1141} demonstrate that, even though tides do not have a significant contribution to periastron precession in PSR J1141-6545 at present, they will be the dominant mechanism after its orbit decays to frequencies $\gtrsim 0.044$ Hz in the next 580 Myr.}

Finally, we perform a similar analysis on PSR B2303+46. As the orbital period of this binary is much longer (see Table \ref{sys_params}), we find that its orbital elements do not evolve significantly due to GR. As a result, PSR B2303+46 will not enter the LISA band within a Hubble time.

Although the two known Galactic NS-WDs with eccentric orbits are not directly relevant to future GW observations and are not candidates for measuring periastron precession, their existence has instigated a number of studies predicting a significant Galactic population of such systems. In the rest of the study we focus on analyzing this predicted population.  

\section{Periastron Precession for the predicted population of eccentric NS-WD binaries} \label{popsynth}

\citet{ty93} and \citet{nyp01} predict a population of galactic NS-WD {binaries} with an evolutionary history similar to that of PSR J1141-6545 and PSR B2303+46. Specifically, they predict $\sim10^4 - 10^6$ systems for the former and $\sim10^6$ systems for the latter. \citet{ty93} and \citet{ts00} provide detailed estimates of the expected distribution of orbital periods, eccentricities and WD masses for such systems (see Table \ref{pop_synth}). Many of these systems will be readily detectable by LISA/eLISA/NGO and tidal effects could potentially leave an imprint on their periastron precession rate. {Thus,} to correctly infer the properties of these systems from periastron precession measurements, it is imperative to understand the role played by tides \citep{wvk08}.

{Apart from eccentric NS-WD binaries that form via binary evolution, \cite{blw13} recently suggested that NS-WD binaries with small but non-zero eccentricities could be formed via three body interactions in globular clusters. We find that all such systems currently known have extremely small eccentricities and thus have a negligible tidal contribution to periastron precession.}

Here we adopt results of population synthesis calculations by \citet{ts00} as they provide the ranges in orbital periods, eccentricities, and WD masses. We analyze the predicted population through time evolution due to GR inspiral and WD cooling with a particular focus on addressing the following questions: (i) what are the binary properties of the systems that evolve into the LISA band? (ii) how strong is the tidal contribution to periastron precession in these systems? (iii) for the subset of systems with a significant tidal contribution, what is the mass overestimate if the sole contribution of GR is taken into account when extracting the WD mass from periastron precession rate measurements?

\begin{deluxetable}{lccrrrr}
\tabletypesize{\scriptsize}
\tablecaption{Parameter Space Predicted by Population Synthesis \label{pop_synth}}
\tablewidth{0pt}
\tablehead{\colhead{Reference} & \colhead{Type of WD} & \colhead{Galactic Population} & \colhead{$M_{_{NS}}(M_{\odot})$} & \colhead{$M_{_{WD}}(M_{\odot})$} & \colhead{$P_{orb}$ (days)} & \colhead{$e$}}
\startdata
\citet{ty93} & CO & $0.14\times10^6 - 0.30\times10^7$ & 1.40 & 0.90 - 1.26 & 0.01 - 1000 & 0.2 - 0.6, 0.8 - 1.0 \\
\citet{ty93} & ONe & $0.99\times10^4 - 0.31\times10^6$ & 1.40 & 1.12 - 1.40 & 0.1 - 1000 & 0.2 - 1.0 \\
\citet{ts00} & not specified & not specified & 1.30 & 0.6 - 1.4 & 0.003 - 1000 & 0.0 - 1.0 \\
\end{deluxetable}

\begin{deluxetable}{lccccc}
\tabletypesize{\scriptsize}
\tablecaption{{Bounds on Orbital Period Correlation to Eccentricity} \label{bound}}
\tablewidth{0pt}
\tablehead{\colhead{Bound} & \colhead{e = 0.2} & \colhead{e = 0.4} & \colhead{e = 0.6} & \colhead{e = 0.8}  & \colhead{e = 1.0}}
\startdata
Lower & 0.0028 days & 0.0046 days & 0.0062 days & 0.0077 days & 0.0090 days \\
Upper & 6.1 days & 15.5 days & 26.7 days & 39.3 days & 1000 days \\
\end{deluxetable}

\subsection{Outline of the Method} 
The computation of the periastron precession rates for the predicted population of eccentric NS-WD binaries proceeds as follows. {First, we scan the parameter space (see Table \ref{pop_synth}) in $M_{_{WD}}$ and $\log_{10}\nu_{orb}$(Hz) in steps of 0.01$M_{\sun}$ and 0.001, respectively.} For each orbital frequency, we use Fig. 4 from \citet{ts00} to derive the eccentricity range to be considered. For instance, systems born with long orbital periods of $\sim$ 1000 days are predicted to have eccentricities of $\sim$ 1. Similarly, systems born with short orbital periods of $\sim$ 0.003 days are predicted to have eccentricities of $\sim$ 0. In Table \ref{bound} we list, for different eccentricities, the orbital period intervals considered, according to Fig. 4 in Tauris and Sennels (2000). Next, we scan on the derived eccentricity range in steps of 0.01. For each combination of the above parameters, we compute the periastron precession rate by summing the contributions of GR, tides, and rotation. 

Once the total periastron precession rate has been calculated, we extract the mass of the WD component assuming that only GR contributes to the precession, thus computing the error in the mass inferred for each combination of WD mass, orbital frequency and eccentricity.

\subsection{Simplifying the Tidal and Rotational Contributions to Periastron Precession: $k_2 R^5$ vs $M_{_{WD}}$} \label{properties}

Following \citet{vfw11}, we investigate whether the tidal and rotational contributions to periastron precession [Eqs. (\ref{tid}) and (\ref{rot}), respectively] can be made dependent solely on the orbital period, eccentricity, and component masses (similarly to the GR contribution in Eq. (\ref{gr})). Specifically, we check whether there is a one-to-one correlation between the term $k_2 R^5$, entering Eqs. (\ref{tid}), and (\ref{rot}), and $M_{_{WD}}$. We use detailed WD models computed with {\tt LPCODE} stellar evolution code \citep{a05}. {\tt LPCODE} has been used to study different problems related to the formation and evolution of WDs: see \citet{a13} and \citet{ram10} for recent applications to the computation of WD cooling sequences. The input physics of the code includes the equation of state of \citet{sc01} for the high-density regime --- which accounts for all the important contributions for both the liquid and solid phases --- complemented with an updated version of the equation of state of \citet{mm79} for the low-density regime. Radiative opacities are those of OPAL \citep{ir96}, including carbon- and oxygen-rich compositions, complemented with the low-temperature opacities of \citet{af94}{\bf. Conductive} opacities are taken from \citet{cpp07}. For effective temperatures less than 10,000 K, outer boundary conditions for the evolving models are  given by detailed non-gray model atmospheres. Recently, {\tt LPCODE} has been tested against other WD evolutionary codes, and uncertainties in the WD cooling ages resulting from different numerical implementations of stellar evolution equations were found to be below 2$\%$ \citep{sag13}.

For our carbon/oxygen core sequences, those with stellar masses less than 1 $M_{\odot}$, the corresponding WD initial configurations are obtained from the full evolution of progenitor stars we computed in previous studies \citep{ram10}. In those studies, progenitor stars were evolved from the zero age main sequence, through the thermally-pulsing and mass-loss phases on the asymptotic giant branch (AGB), to the WD cooling phase. For our more massive WD models, those with M $>$ 1.06 $M_{\odot}$, we have assumed a core composition of oxygen/neon (see \citet{a07} for details).

\begin{figure}
\epsscale{1.17}
\plottwo{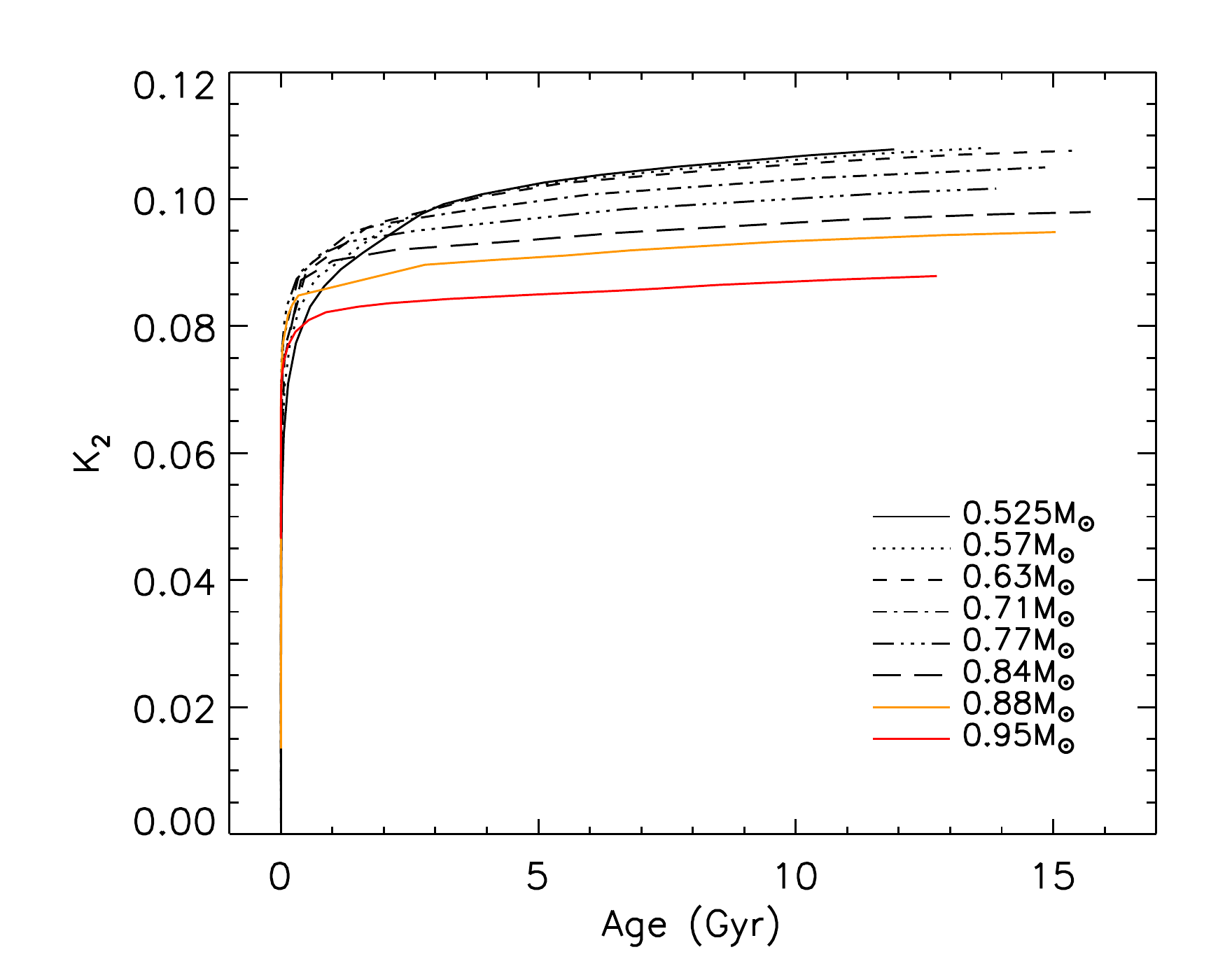}{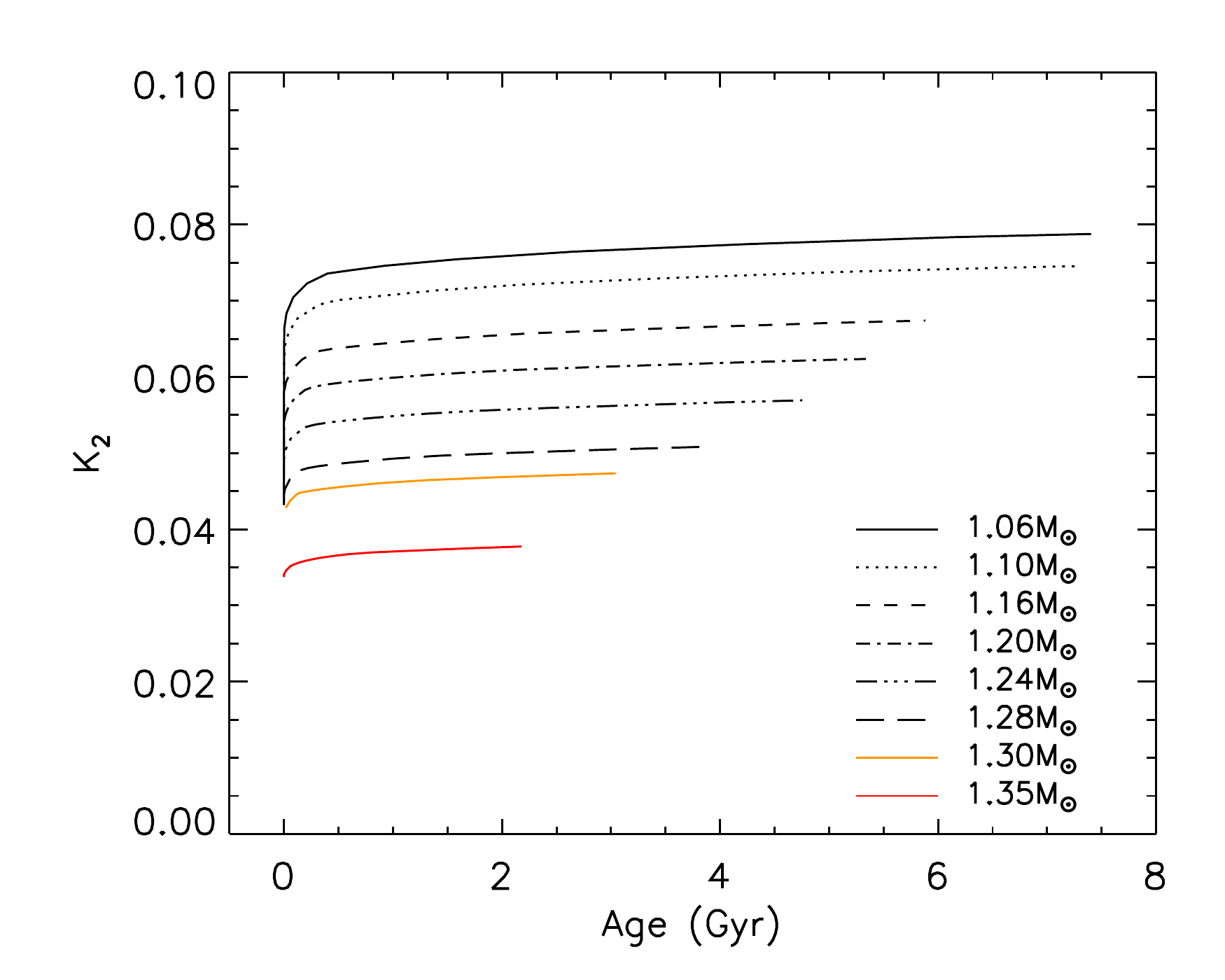}
\caption{$k_2$ as a function of age for CO (left) and ONeMg (right) WDs.}
 \label{K2}
\end{figure}

\begin{figure}
\epsscale{1.15}
\plottwo{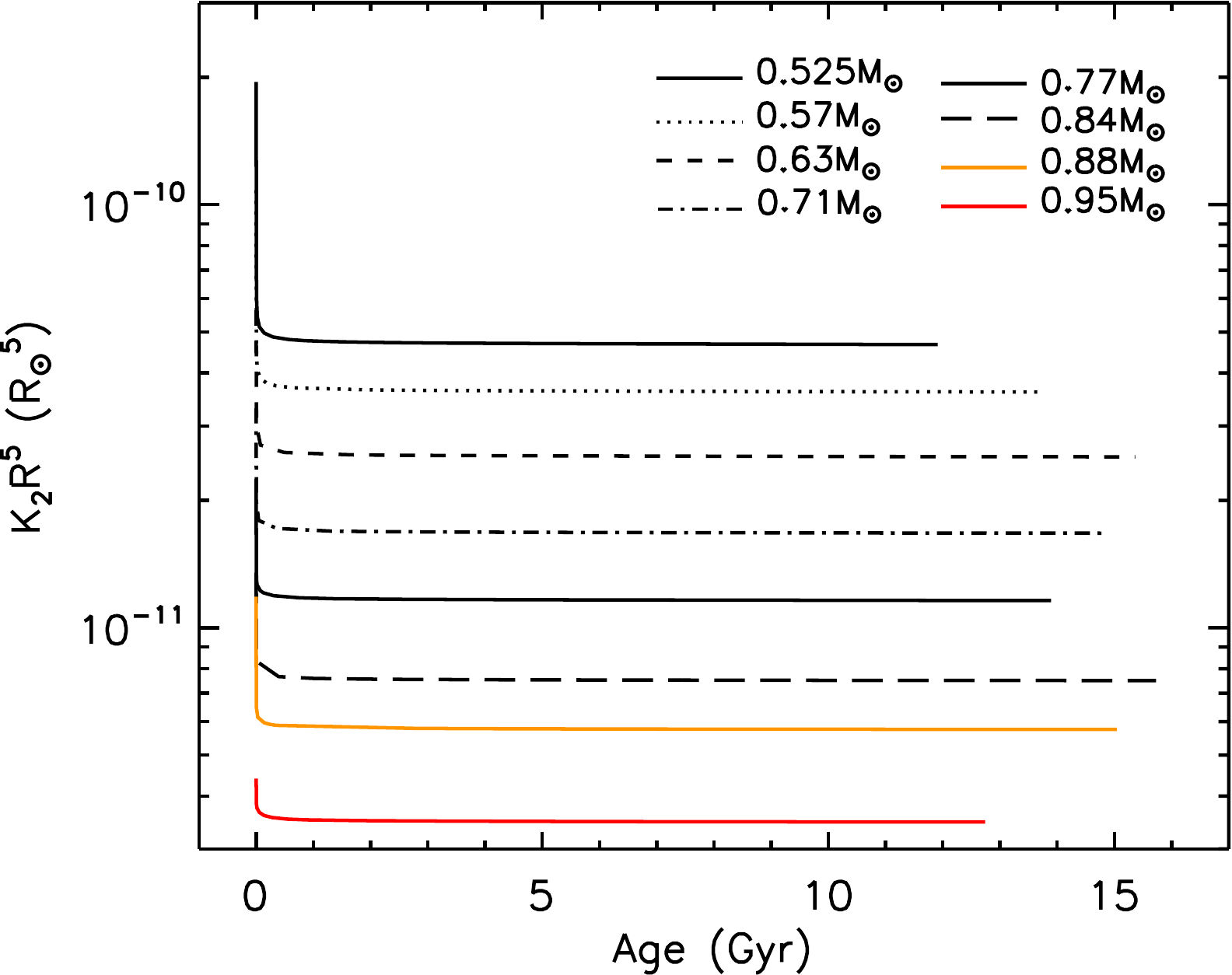}{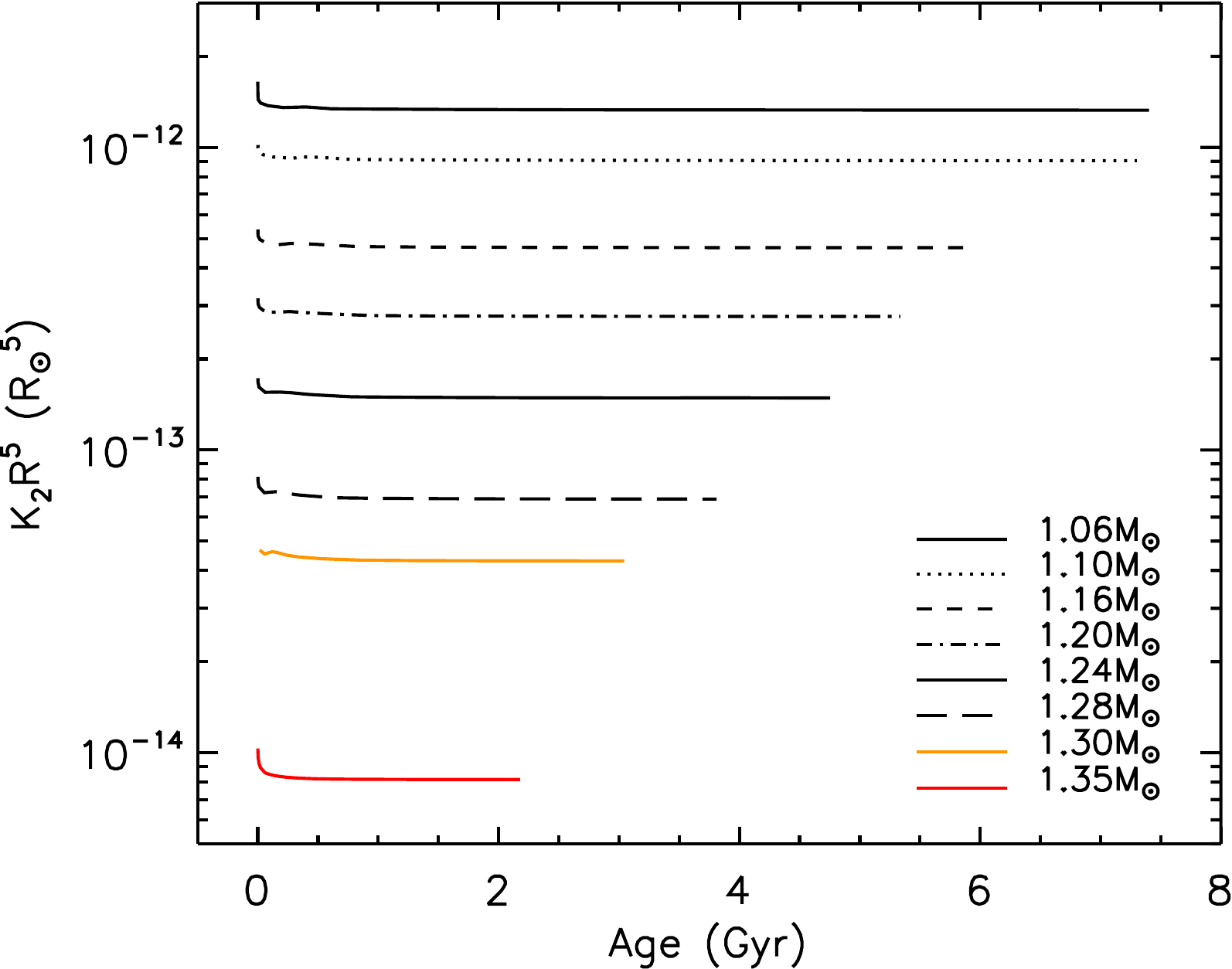}
\caption{$k_2R^5$ as a function of age for CO (left) and ONeMg (right) WDs.}
 \label{KR5}
\end{figure}

To calculate $k_2$, we use the WD models described above and Eqs. (\ref{k2}), (\ref{de}) and (\ref{bc}). The radius of the WD is taken from the model. In Figs. \ref{K2} and \ref{KR5} we plot $k_2$ and $k_2R^5$ as a function of the WD age, respectively, for different WD models. Fig. \ref{K2} shows that, for each WD model, $k_2$ increases as the WD evolves. This means that the star becomes less centrally concentrated as it cools. Fig. \ref{KR5} shows that after 0.5 Gyr of cooling, all WDs reach a plateau value of $k_2R^5$, which is held nearly constant for the stars' remaining lifetime. We also note that the values of $k_2R^5$ vary by less than an order of magnitude over the lifetime of any WD and the variation becomes less significant with increasing WD mass. {This behavior explains the effect described in \S~\ref{observed}, where we find that the periastron precession rates for PSR J1141-6545 are not significantly affected when WDs of different ages are considered.}

\begin{figure}
\epsscale{0.65}
\plotone{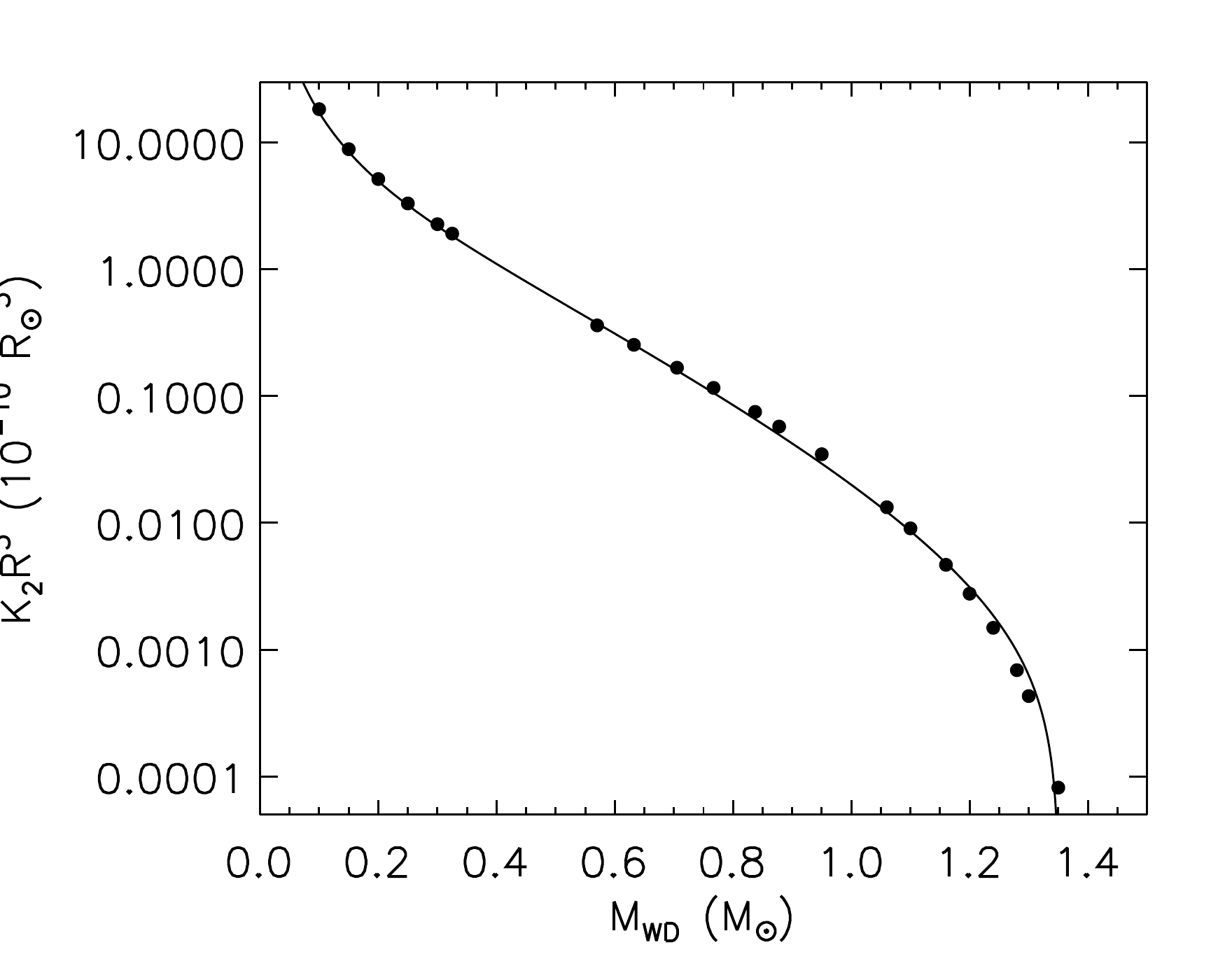}
\caption{$k_2R^5$ as a function of $M_{_{WD}}$ for WDs $\geq$ 0.5 Gyr old.}
 \label{corres}
\end{figure}

In Fig. \ref{corres} we plot plateau values of $k_2R^5$ as a function of $M_{_{WD}}$ for the WD models described above and those described in \citet{vfw11} and plotted in Fig. 4 therein (seen here as a series of 6 detached points between 0.1 $M_\odot$ and 0.35 $M_\odot$). We can see that the values of $k_2R^5$ vary smoothly with $M_{_{WD}}$ for the entire set of WD {models}. This trend is described well by the following fitting formula:

\begin{equation} \label{fit}
k_2R^5~(10^{-10}R_\odot^5) = 0.73~M_{_{WD}}^{-\frac{3}{2}}~exp(-3.6~M_{_{WD}}) - 8.0~exp(-4.8~M_{_{WD}})~log(M_{_{WD}})~sin(M_{_{WD}})
\end{equation}

where $M_{_{WD}}$ is in solar units. This relation solves the degeneracy problem indentified by \citet{wvk08} and mentioned in \S~\ref{Intro} by eliminating the dependence of Eqs. (\ref{tid}) and (\ref{rot}) on the WD central concentration and radius. Therefore, if the orbital period and eccentricity of a NS-WD binary (older than 0.5 Gyr) are known (from e.g. the frequency spectrum of the GW signal from these sources), one can extract the mass of the WD component from periastron precession rate measurements. Since Eq. (\ref{fit}) is only satisfied by binaries older than 0.5 Gyr, in the rest of our analysis, we only study the population of systems $\geq$ 0.5 Gyr old.

\subsection{Evolving the Parameter Space Forward in Time}

\begin{figure}
\epsscale{1.22}
\plotone{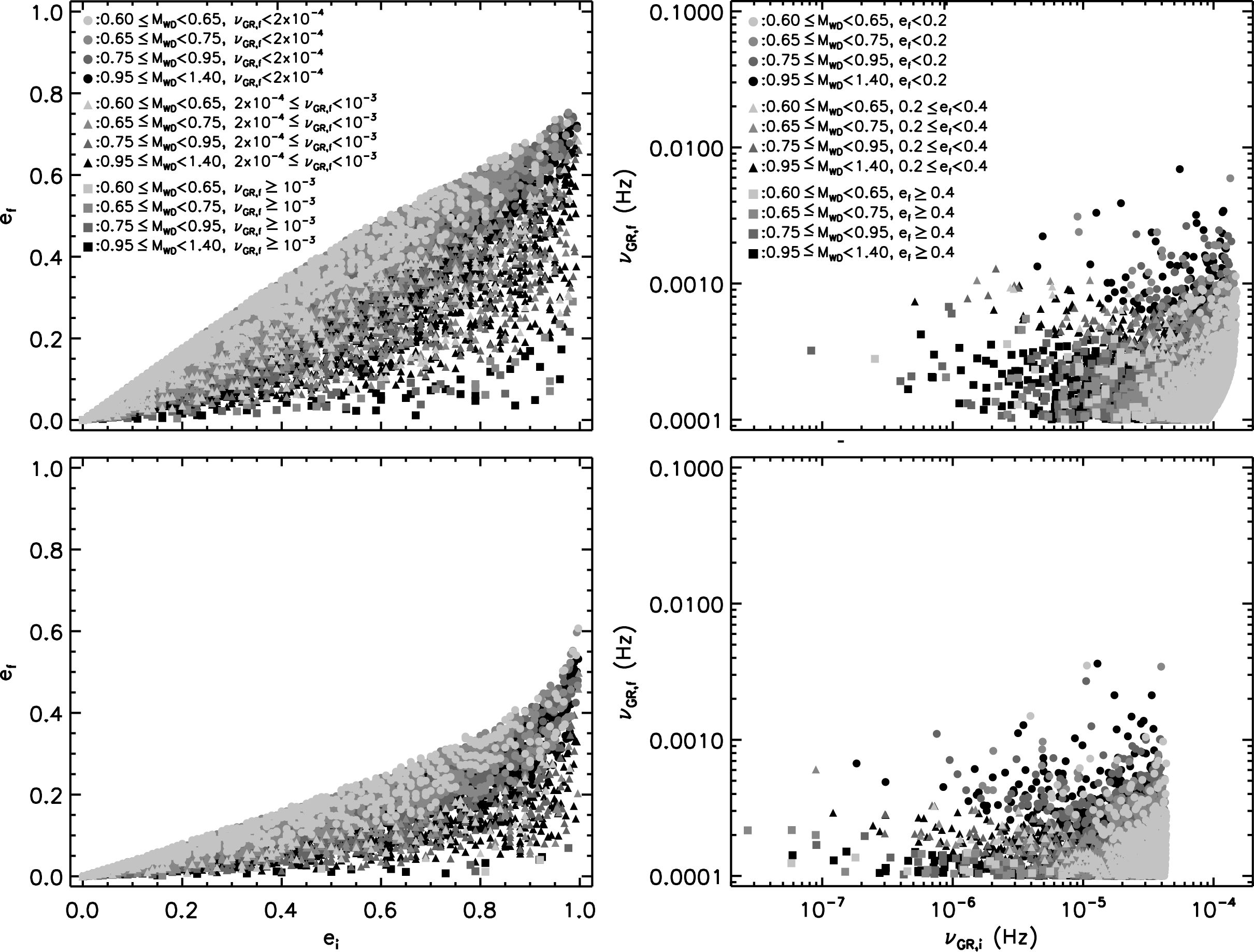}, 
\caption{{Final vs. initial eccentricity (left), final vs. initial GW frequency (right) for WDs that are 0.5 Gyr old (top) and 13.8 Gyr old (bottom). The WD mass and frequency are in $M_{\odot}$ and Hz, respectively. The cut-off seen in the right panels show that systems with initial GW frequencies higher than $\sim 2\times 10^{-4} ~( \sim5\times 10^{-5})$ Hz will evolve outside of the LISA band within 0.5 (13.8) Gyr. We note that the GW frequency at which a cold $1.35M_\sun$ ($0.6M_\sun$) WD would overfill its Roche lobe is $0.48$ Hz ($0.034$ Hz). Since the GW frequency of a system increases with time due to GR, the upper limits on the final GW frequency seen here show that none of the systems underwent or {are} undergoing a phase of Roche lobe overflow.}}
\label{0.5and13.8evol}
\end{figure}

\begin{figure}
\epsscale{1.17}
\plotone{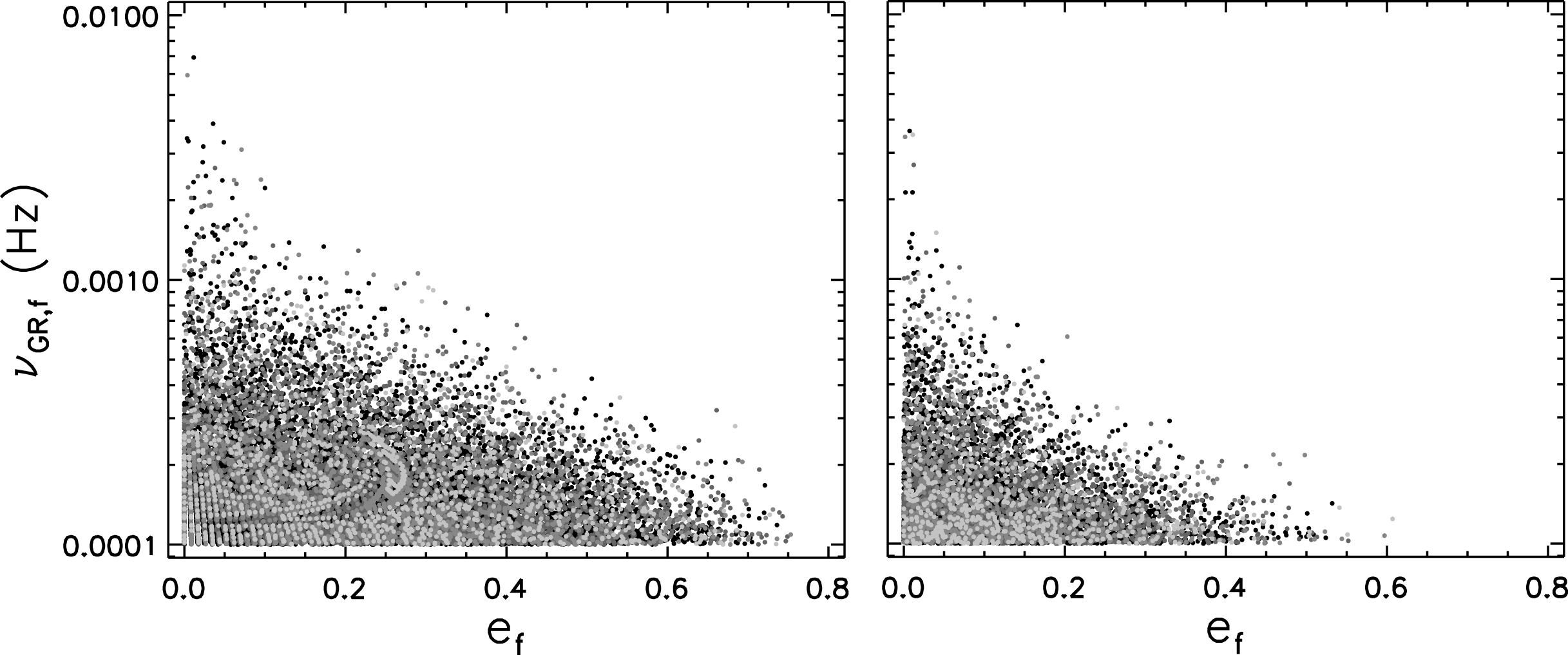}
\caption{{Final GW frequency vs. final eccentricity of systems predicted by population synthesis models that have evolved for 0.5 Gyr (left) and 13.8 Gyr (right). The color scheme used is the same as in Fig. \ref{0.5and13.8evol}.}}
\label{nuFinal_eFinal}
\end{figure}

Given the behavior of $k_2R^5$ presented in \S~\ref{properties}, here we study the effect of tides on periastron precession for systems older than 0.5 Gyr. In particular, we consider two evolutionary stages: 0.5 Gyr and 13.8 Gyr corresponding to the age at which the component WD attains its plateau value of $k_2R^5$ (described by Eq. (\ref{fit})) and equal to a Hubble time, respectively. As the orbit of {these} binaries will evolve due to GR, it is important to determine how the parameter space predicted by population synthesis calculations and described in the previous sections changes after each system is evolved for 0.5 {and 13.8 Gyr}. Additionally, the sensitivity of LISA will place an upper (a lower) limit on the {orbital frequency (orbital period)}.

{In the top (bottom) panels of Fig. \ref{0.5and13.8evol} we plot the initial and final eccentricity ($e_{i}$ and $e_{f}$, respectively), and the initial and final GW frequency ($\nu_{GR,i}$ and $\nu_{GR,f}$, respectively) for 0.5 Gyr (13.8 Gyr) old systems. In the left (right) panel of Fig. \ref{nuFinal_eFinal}, we show the final GW frequency and eccentricity for 0.5 Gyr (13.8 Gyr) old systems. From Fig. \ref{nuFinal_eFinal}, we see that evolved systems with higher GW frequencies are associated with smaller eccentricities and vice versa.}

From Eqs. (\ref{tid}), (\ref{rot}), and (\ref{gr}) we can see that the tidal, rotational, and GR contributions to periastron precession have different dependences on the components' masses and properties. As noted in \S~\ref{Intro}, assuming the sole contribution of GR when extracting the component's masses from periastron precession rate measurements could lead to overestimating the mass. In this study, our method to determine the mass bias if only GR is accounted for proceeds as follows. First, we compute the periastron precession rate for each binary configuration, considering the sum of the contributions of GR, rotation, and tides. Then, we extract the mass of the WD component assuming that GR is the only driver of periastron precession.

In the next section, we use the evolved parameter spaces at 0.5 and 13.8 Gyr, shown in Fig. \ref{nuFinal_eFinal}, to investigate the importance of tides in driving periastron precession in NS-WD binaries.

\subsection{The Periastron Precession Rates} \label{results}

\begin{figure}
\epsscale{0.9}
\plotone{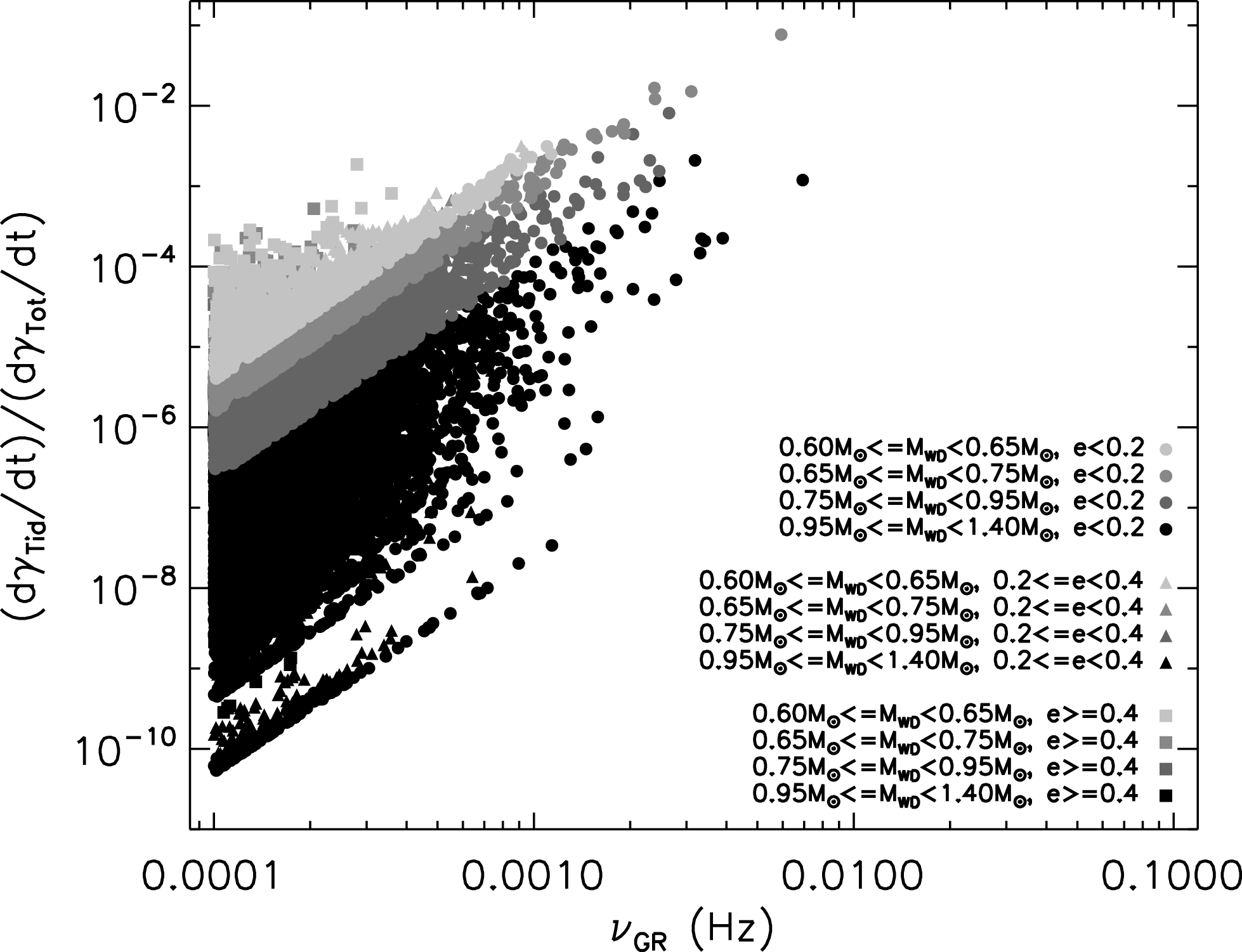}
\caption{{Ratio of the periastron precession rate due to tides to the total periastron precession rate ($\dot{\gamma}_{Tid}/\dot{\gamma}_{tot}$) vs. GW frequency ($\nu_{GR}$) for systems that are 0.5 Gyr old. $M_{_{WD}}$ is in units of $M_{\odot}$.}}
 \label{plot1_05}
\end{figure}

\begin{figure}
\epsscale{0.9}
\plotone{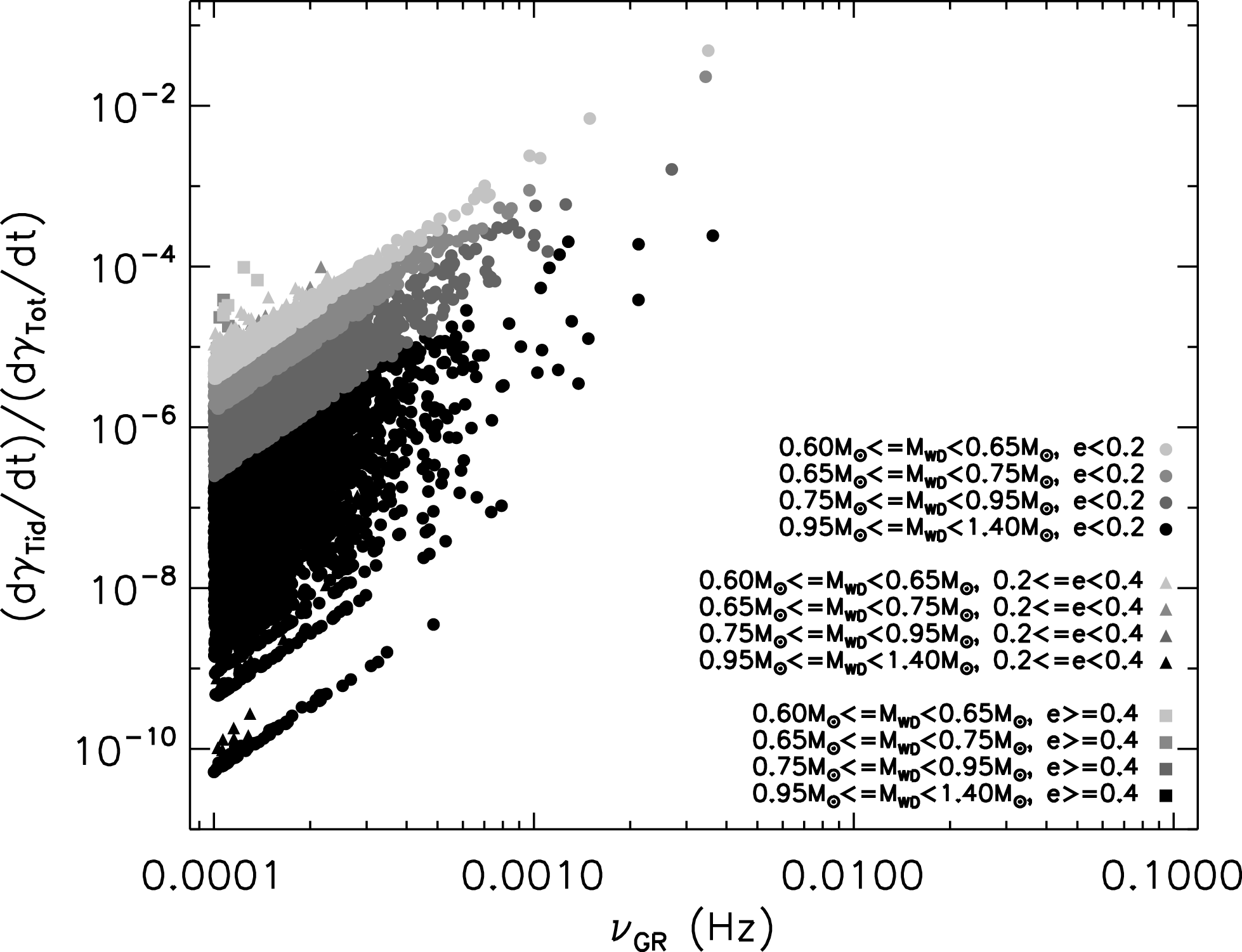}
\caption{{Ratio of the periastron precession rate due to tides to the total periastron precession rate ($\dot{\gamma}_{Tid}/\dot{\gamma}_{tot}$) vs. GW frequency ($\nu_{GR}$) for systems that are 13.8 Gyr old. $M_{_{WD}}$ is in units of $M_{\odot}$.}}
 \label{plot1_13.8Gyr}
\end{figure}

\begin{figure}
\epsscale{1.17}
\plotone{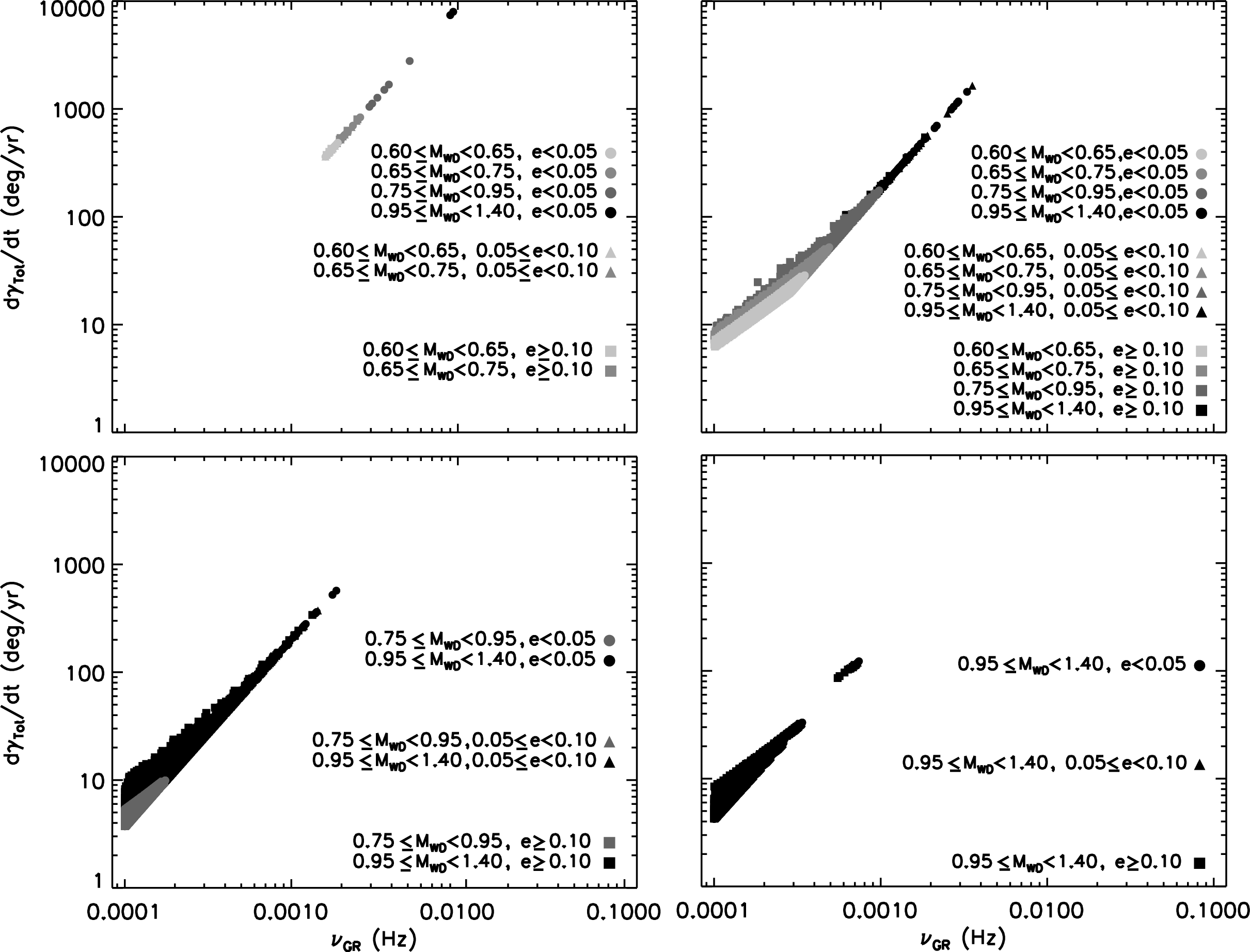}
\caption{{$\dot{\gamma}$ as a function of $\nu_{GR}$ for $\dot{\gamma}_{Tid}/\dot{\gamma}_{tot} \sim$ $10^{-2}$ (top-left), $10^{-4}$ (top-right), $10^{-6}$ (bottom-left), and $10^{-8}$ (bottom-right) in 0.5 Gyr old systems.  $M_{_{WD}}$ is in units of $M_{\odot}$.}}
 \label{rat05}
\end{figure}

\begin{figure}
\epsscale{1.17}
\plotone{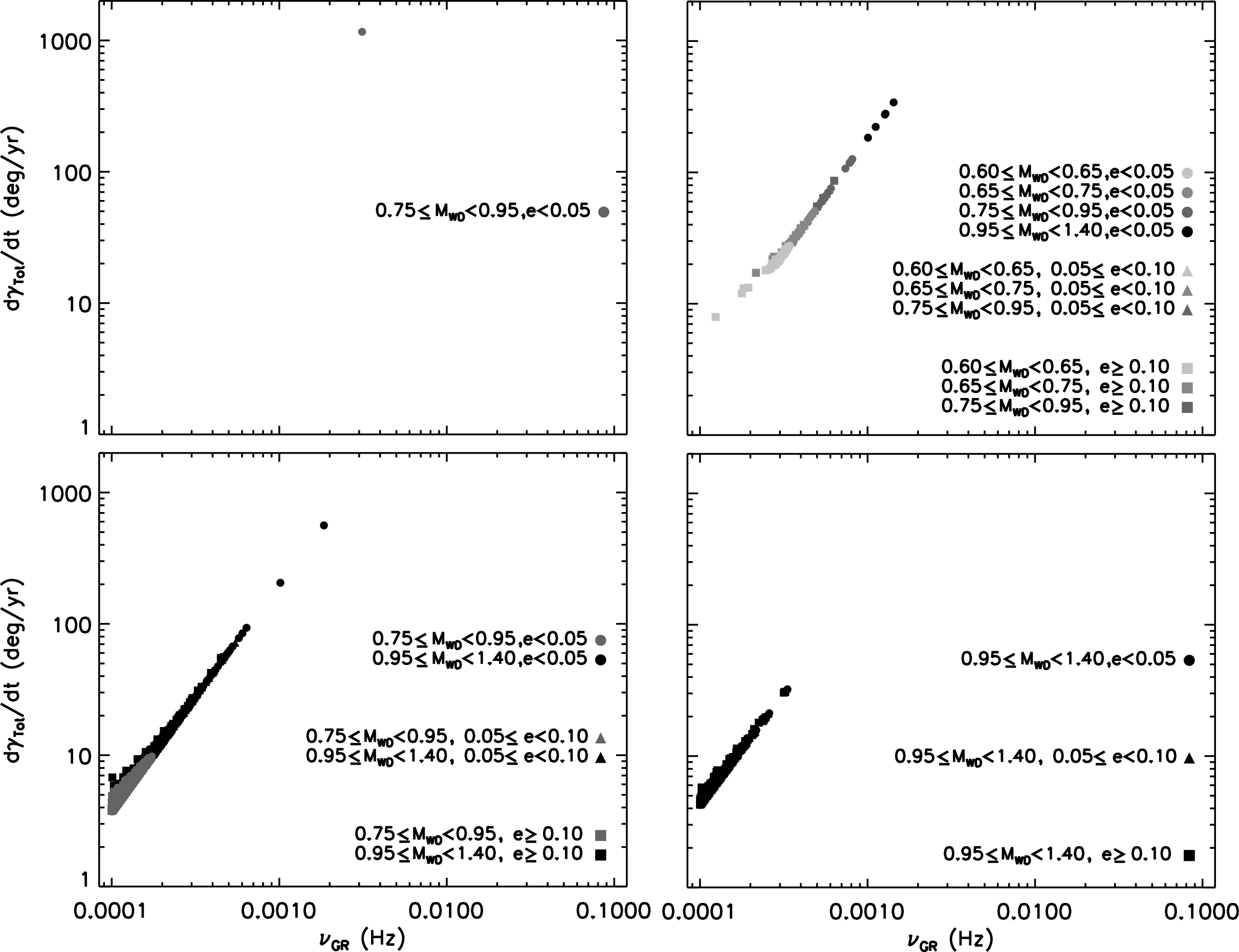}
\caption{{$\dot{\gamma}$ as a function of $\nu_{GR}$ for $\dot{\gamma}_{Tid}/\dot{\gamma}_{tot} \sim$ $10^{-2}$ (top-left), $10^{-4}$ (top-right), $10^{-6}$ (bottom-left), and $10^{-8}$ (bottom-right) in 13.8 Gyr old systems.  $M_{_{WD}}$ is in units of $M_{\odot}$.}}
 \label{rat138}
\end{figure}

\begin{figure}
\epsscale{0.9}
\plotone{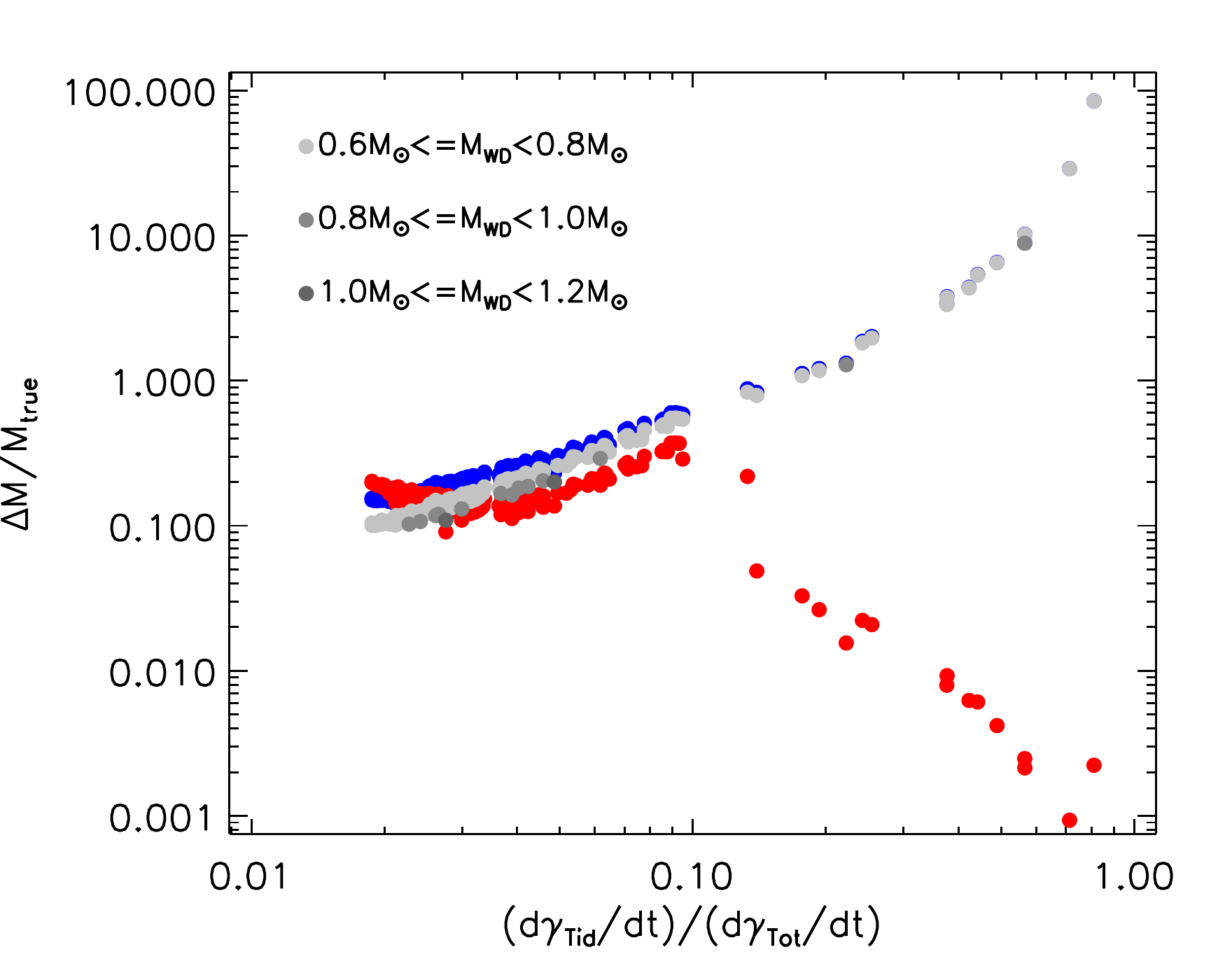}
\caption{Relative error in the WD mass estimate as a function of $\dot{\gamma}_{Tid}/\dot{\gamma}_{tot}$ for 0.5 Gyr old systems due to neglecting tides (in grey), the NS mass uncertainty (in red), and both neglecting tides and the NS mass uncertainty (in blue).}
 \label{shift05}
\end{figure}

\begin{figure}
\epsscale{0.9}
\plotone{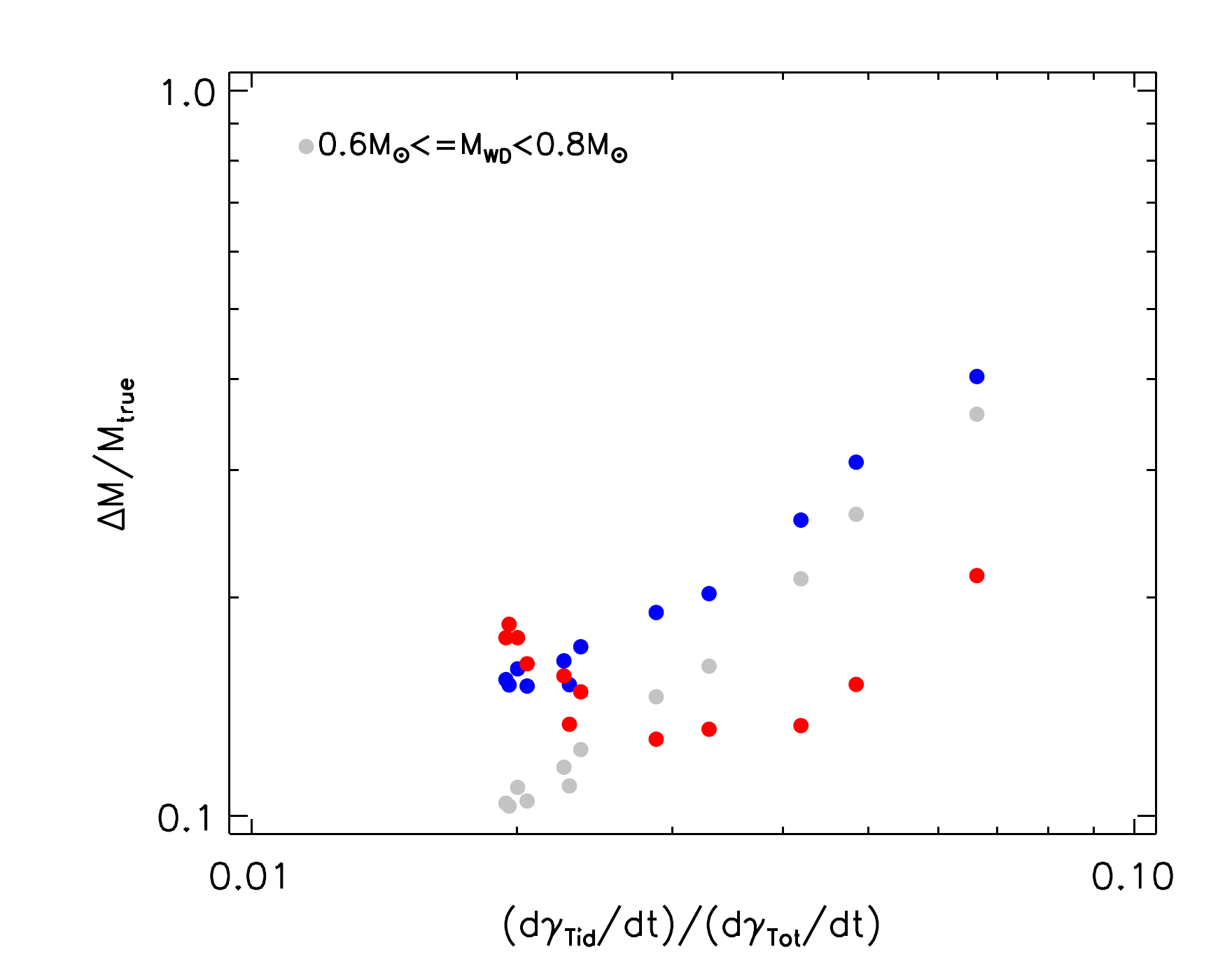}
\caption{Relative error in the WD mass estimate as a function of $\dot{\gamma}_{Tid}/\dot{\gamma}_{tot}$ for 13.8 Gyr old systems due to neglecting tides (in grey), the NS mass uncertainty (in red), and both neglecting tides and the NS mass uncertainty (in blue).}
 \label{shift138}
\end{figure}

In Fig. \ref{plot1_05} we plot the ratio of periastron precession rate due to tides to the total periastron precession rate ($\dot{\gamma}_{Tid}/\dot{\gamma}_{tot}$) as a function of the {GW} frequency ($\nu_{GR}$) for 0.5 Gyr old systems in 3 eccentricity ranges: $e < 0.25$, $0.25 \leq e < 0.5$, and $e \geq 0.5$. The total periastron precession rate accounts for GR, tides, and rotation. Fig. \ref{plot1_13.8Gyr} shows a similar calculation for 13.8 Gyr old systems. We observe that systems with heavier WDs (darker points) have lower values of $\dot{\gamma}_{Tid}/\dot{\gamma}_{tot}$ than systems with lighter WDs (lighter points). This is because heavier WDs have smaller values of $k_2R^5$ than lighter WDs (see \S~\ref{properties}), resulting in a lower tidal contribution to the periastron precession rate. In Fig. \ref{rat05} we plot the total periastron precession rate as a function of $\nu_{GR}$ for 0.5 Gyr systems where $\dot{\gamma}_{Tid}/\dot{\gamma}_{tot}$ is $\sim$ $10^{-2}$ (top-left), $10^{-4}$ (top-right), $10^{-6}$ (bottom-left), and $10^{-8}$ (bottom-right). Fig. \ref{rat138} shows similar results for 13.8 Gyr systems. As expected, the bottom-right panels in both figures are populated by heavy WDs (dark points). Similarly, the converse is seen {in} the top-left panels, where we can also see lighter WDs (lighter points). The value of the total periastron precession rate can be as high as $\sim 10^{4}$ deg/yr (as seen from the top-left panel of Fig. \ref{rat05}). Additionally, systems with a higher $\dot{\gamma}_{Tid}/\dot{\gamma}_{tot}$ also have a higher {$\dot{\gamma}_{tot}$}.

{So far, all our calculations assumed that the NS component has a mass of 1.3 $M_{\odot}$, following population synthesis models by \citet{ts00}. However, NSs in eccentric NS-WD binaries are observed with masses from 1.27 $M_{\odot}$ to 1.40 $M_{\odot}$ (see Table \ref{sys_params}) and, when interpreting GW measurements, the actual NS mass might be unknown. Here we quantify how the uncertainty in the NS mass affects the WD mass inferred from periastron precession measurements in two ways: (i) we isolate the error introduced by the NS mass uncertainty and (ii) we compute the worst case error, arising from both neglecting tides and the NS mass uncertainty.

Figs. \ref{shift05} and \ref{shift138} show the different errors entering the determination of the WD component mass for 0.5 Gyr and 13.8 Gyr old systems, respectively. In grey we plot the error arising from neglecting tides and we only show systems for which the relative error is $\geq 10\%$. Then, for each grey data point, we calculate the total periastron precession rate. Given $\dot{\gamma}_{tot}$,  we fix the orbital parameters, we vary the NS mass between 1.27 $M_{\odot}$ and 1.40 $M_{\odot}$, and we extract two values for the WD mass, $M_{_{WD}}^{NS_1}$ and $M_{_{WD}}^{NS_2}$. The error due to the uncertainty in $M_{_{NS}}$ alone is then evaluated as

\begin{equation} \label{err1}
\left(\frac{\Delta M}{M_{true}}\right)_{NSmass} = \frac{M_{_{WD}}^{NS_1} - M_{_{WD}}^{NS_2}}{M_{_{WD}}^{NS_{true}}},
\end{equation}
where $M_{_{WD}}^{NS_{true}}$ is the WD mass computed for each grey data point in Figs. \ref{shift05} and \ref{shift138} assuming a NS mass of 1.3 $M_{\odot}$. For systems that yield two solutions for $M_{_{WD}}^{NS_{(1,2)}}$ (see Appendix), the maximum value of Eq. (\ref{err1}) is used. The uncertainties thus computed are shown in red in  Figs. \ref{shift05} and \ref{shift138}. 
 
Finally, we compute the worst case error arising from both neglecting tides and the NS mass uncertainty. As before, we calculate $\dot{\gamma}_{tot}$ for each grey data point in Figs. \ref{shift05} and \ref{shift138}.  We then set $M_{_{NS}} =$ 1.27 $M_{\odot}$ and 1.40 $M_{\odot}$, and extract two WD masses, $M_{_{WD}}^{1.27}$ and $M_{_{WD}}^{1.40}$, respectively, assuming the sole contribution of GR. The worst case error is then evaluated as

\begin{equation} \label{err2}
\left(\frac{\Delta M}{M_{true}}\right)_{NSmass,tides} = Max \left[\frac{(M_{_{WD}}^{1.27} - M_{_{WD}}^{NS_{true}})}{M_{_{WD}}^{NS_{true}}}~, ~\frac{(M_{_{WD}}^{1.40} - M_{_{WD}}^{NS_{true}})}{M_{_{WD}}^{NS_{true}}}\right].
\end{equation}
The error thus computed is shown in blue in Figs. \ref{shift05} and \ref{shift138}.

Comparing Fig. \ref{shift05} (Fig. \ref{shift138}) with Fig. \ref{plot1_05} (Fig. \ref{plot1_13.8Gyr}) we find that systems that incur an error of $>$ 10\% due to ignoring tides have higher orbital frequencies. Since their orbits decay rapidly through this regime (for example, the orbital decay time, $t_{GR} = |a/\dot{a}|$, for a circular binary with $M_{_{NS}} =$ 1.3 $M_{\odot}$, $M_{_{WD}} =$ 0.6 $M_{\odot}$, and $\nu_{GR} = 0.01$ Hz is $\sim 28000$ years), they constitute a small fraction of the predicted population. We note that these systems should also be easier to detect owing to their relatively large GW strains. 

The small fraction of systems populating Figs. \ref{shift05} and \ref{shift138} have $\dot{\gamma}_{Tid}/\dot{\gamma}_{tot} >$ 0.01 and can incur errors of $>$ 10\% due to ignoring tides. However, in systems where 0.01 $< \dot{\gamma}_{Tid}/\dot{\gamma}_{tot} \lesssim$ 0.03, we find that the NS mass uncertainty dominates the WD mass uncertainty. Whereas, in systems where $\dot{\gamma}_{Tid}/\dot{\gamma}_{tot} >$ 0.03, the dominant uncertainty arises due to ignoring tides. Additionally, when $\dot{\gamma}_{Tid}/\dot{\gamma}_{tot} \gtrsim$ 0.1, the error in the WD mass is $>$ 90\% and can be as high as $\sim$ 8000\%, potentially leading to a misclassification of the source, if tides are neglected. Thus, if $\dot{\gamma}_{Tid}/\dot{\gamma}_{tot} \gtrsim$ 0.03, Eq. (\ref{fit}) can be used to place constraints on the WD mass estimated from periastron precession rate measurements.

Finally, we find that the error due to the NS mass uncertainty is always $<$ 40\% ($\lesssim$ 20\%) in 0.5 (13.8) Gyr old systems, and that it decreases rapidly to $<$ 0.1\% for systems with $\dot{\gamma}_{Tid}/\dot{\gamma}_{tot} \gtrsim$ 0.1, if tides are properly accounted for. We note that the errors become flat at small $\dot{\gamma}_{Tid}/\dot{\gamma}_{tot}$ values. As explained in the Appendix, this occurs because the range of NS masses that yield a solution for the WD mass becomes narrower as the relative tidal contribution decreases. This implies that periastron precession rate measurements can also be used to constrain the NS component mass in systems with $\dot{\gamma}_{Tid}/\dot{\gamma}_{tot} <$ 0.1.}
 
\section{Conclusions} \label{conclusions}
GW emission from eccentric binaries encodes information about the rate at which the periastron of their orbit precesses. The periastron precession rate depends on the components' properties and orbital parameters. Therefore, periastron precession measurements can be used to constrain some of the binary properties. {In this work, we focus on such precession in eccentric NS-WD binaries. These GW sources are of particular interest because their periastron precession rate can be used to place constraints on their WD component's mass if the orbital parameters are known. Additionally, since pulsar-timing measurements could yield the components' masses independent of precession effects, these systems can be used to verify the validity of our models. Here we investigate the significance of the three mechanisms driving periastron precession, namely tides, rotation, and GR. In particular, we focus on the tidal contribution and investigate the errors introduced in the WD mass estimated from periastron precession rate measurements, if tides are ignored.} 

First, we analyze tides in the two eccentric NS-WD binaries currently known: PSR J1141-6545 and PSR B2303+46. These are believed to have formed via a common envelope phase followed by a supernova event that formed the NS. For both systems, we find that tides are not the dominant driver of periastron precession at present. However, for the case of J1141-6545, we find that, as the system evolves due to GR, tides will take over as the leading mechanism in 580 Myr.   

Even though we observe only two NS-WD binaries with a significant eccentricity, population synthesis studies predict the existence of a host of such systems in our galaxy with a wide range of orbital periods and eccentricities. {We analyze periastron precession in the predicted sources and find that the tidal contribution is stronger in systems having high orbital frequencies, high eccentricities, and low WD masses (i.e., larger radii). Furthermore, the tidal contribution grows more rapidly with increasing orbital frequency than eccentricity. In majority of the systems the relative tidal contribution is small ($<$ 1\%) and the error in the WD mass inferred would be $<$ 10\% if tides are ignored. In systems where the relative tidal contribution is between 1\% and 3\%, the NS mass uncertainty dominates the errors in the WD mass inferred. In systems where the relative tidal contribution is between 3\% and 10\%, the errors arising due to ignoring tides dominate. However, in this regime, the NS mass uncertainty limits the accuracy with which the WD mass can be inferred. Regardless, the error arising solely due to the NS mass uncertainty is always lesser than 40\% (20\% for the oldest systems). Finally, in systems where the relative tidal contribution is $>$ 10\%, tides play a significant role and, if neglected, the error in the WD mass inferred from periastron precession measurements is $>$ 90\% and could be as high as $\sim$ 8000\%. Clearly, in these extreme cases, neglecting tides would lead to a misclassification of the source. However, since systems with relative tidal contribution $>$ 1\% have higher orbital frequencies, they decay rapidly, thereby constituting only a small fraction of the predicted population.

Accounting for the tidal (and rotational) contributions to periastron precession introduces as additional parameters the periastron precession constant $k_2$ and the WD radius R, as $k_{2}R^{5}$. We show that $k_{2}R^{5}$ is a smooth function of the WD mass for most of the WD lifetime (starting at 0.5 Gyr of its cooling age) and derive a relation between $k_{2}R^{5}$ and the WD mass. This relation can be used to simplify the equations governing the total periastron precession rate, facilitating a more accurate extraction of the WD mass from periastron precession measurements.

We conclude that while accounting for tides when interpreting periastron precession rates to determine the WD component's mass in eccentric NS-WD binaries is not necessary in most cases, tidal precession can be accounted for by using the relation between $k_{2}R^{5}$ and WD mass presented here, thereby improving constraints on inferred WD mass in some cases. Gravitational waves emerging from such sources will provide a new astrophysical laboratory to test the reliability of our current understanding of the physics governing these systems and perhaps even unravel elements of the engines fueling compact object physics that remain shrouded by conventional observation techniques in the electromagnetic spectrum.}

\begin{acknowledgements}
{We thank Bart Willems for useful discussions, and the anonymous referee for his/her positive and constructive review. Vicky Kalogera is grateful for support through a Simons Foundation Fellowship in Theoretical Physics and for the hospitality of the Aspen Center for Physics.}
\end{acknowledgements}

\appendix

\section{Constraining the Component Masses From Periastron Precession Rates}

Here we explain the procedure proposed in \S~\ref{results} to compute the error in the extracted WD mass arising from the NS mass uncertainty alone. First, for a given system, the total periastron precession rate is computed using Eqs. (\ref{tid}), (\ref{rot}), and (\ref{gr}), assuming a NS mass of 1.3 $M_{\odot}$. Next, we vary the NS mass between 1.27 $M_{\odot}$ and 1.40 $M_{\odot}$, while keeping the orbital parameters fixed and, for each NS mass, we search for values of $M_{_{WD}}$ that yield the same $\dot{\gamma}_{tot}$. The computation of these values are not straightforward and warrants a discussion of the behavior of $\dot{\gamma}_{tot}$ as a function of $M_{_{WD}}$. Since this behavior changes significantly as $\dot{\gamma}_{Tid}/\dot{\gamma}_{tot}$ increases, we focus on the two systems at the extreme ends of Fig. \ref{shift05}, i.e. the ones with the smallest (A) and largest (B) value of $\dot{\gamma}_{Tid}/\dot{\gamma}_{tot}$. Their properties are outlined in Table \ref{NSmasstestsystems}. 

\begin{deluxetable}{lrrrrrrr}
\tabletypesize{\scriptsize}
\tablecaption{{Properties of System A and B} \label{NSmasstestsystems}}
\tablewidth{0pt}
\tablehead{\colhead{System} & \colhead{$M_{_{NS}}(M_{\odot})$} & \colhead{$M_{_{WD}}(M_{\odot})$} & \colhead{$e$} & \colhead{$\nu_{orb}$} & \colhead{$\dot{\gamma}_{tot}$ (deg/yr)} & \colhead{$\dot{\gamma}_{Tid}/\dot{\gamma}_{tot}$} & \colhead{$\Delta M/M_{true}$}}
\startdata
A & 1.30 & 0.60 & 1.00$\times10^{-1}$ & 1.03638$\times10^{-3}$ & 5.643$\times10^{2}$ & 0.018744 & 0.10435 \\
B & 1.30 & 0.60 & 1.43$\times10^{-3}$ & 9.88927$\times10^{-3}$ & 2.147$\times10^{5}$ & 0.811203 & 84.4638 \\
\enddata
\end{deluxetable}

\subsection{System A}

\begin{figure}
\epsscale{1.17}
\plotone{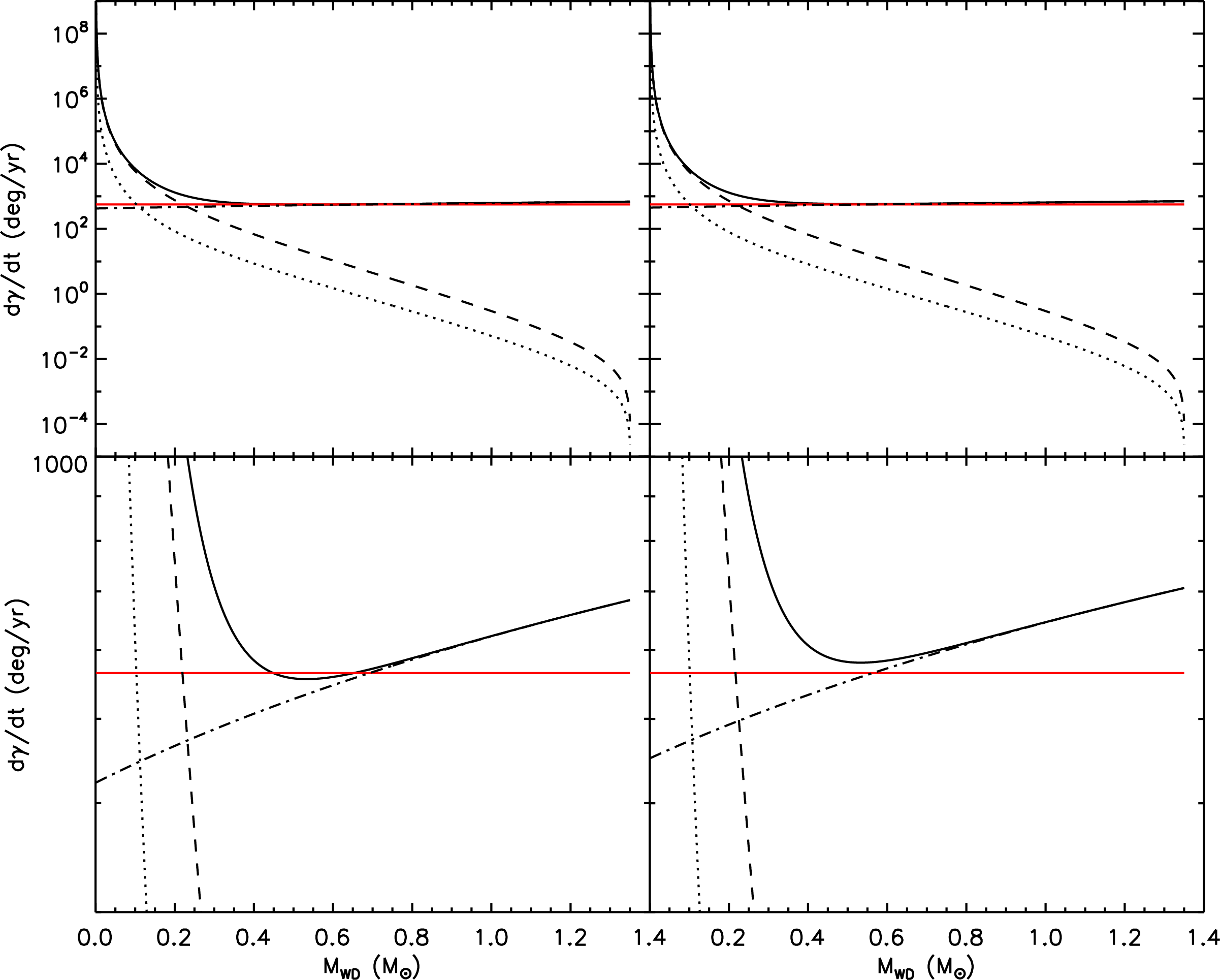}
\caption{System A: $\dot{\gamma}$ as a function of $M_{_{WD}}$ for $M_{_{NS}}$ changed to 1.27 $M_{\odot}$ (left) and 1.40 $M_{\odot}$ (right). The dashed line indicates the tidal contribution, the dotted line indicates the rotational contribution and the dot-dashed line indicates the GR contribution. The solid line represents the total periastron precession rate. The solid red line is the total periastron precession rate we want to match by changing the WD mass. The bottom panels show a blow-up of the regions where the solid red and black lines are comparable in magnitude.}
 \label{sysA}
\end{figure}

{This system has the smallest value of $\dot{\gamma}_{Tid}/\dot{\gamma}_{tot}$ that yields an error $>$10\% in the WD mass estimated from periastron precession rate measurements due to ignoring tides. In Fig. \ref{sysA} we plot the individual and total contributions to periastron precession rate as a function of $M_{_{WD}}$ on changing the NS mass in this system to 1.27 $M_{\odot}$ (left) and 1.40 $M_{\odot}$ (right) in black. The solid red line shows the value of the total periastron precession rate we want to match (see Table \ref{NSmasstestsystems}). We search for values of $M_{_{WD}}$ where the solid black and red curves intersect. We find that the tidal contribution to periastron precession is stronger for smaller values of $M_{_{WD}}$. The reverse is seen for the GR contribution. Therefore, the total periastron precession rate first decreases and then increases as a function of $M_{_{WD}}$. The bottom panels of Fig. \ref{sysA} show a blow-up of the region where the two curves are comparable in magnitude for each $M_{_{NS}}$. We find that the total periastron precession rate increases on increasing $M_{_{NS}}$. This results in two solutions for $M_{_{WD}}$ when $M_{_{NS}}$ is set to 1.27 $M_{\odot}$ and no solutions when set to 1.40 $M_{\odot}$.  This behavior is typical for all systems where $\dot{\gamma}_{Tid}/\dot{\gamma}_{tot} \lesssim$ 0.1. Therefore, to find a solution to Eq. (\ref{err1}) for a system like System A, $M_{_{NS}}$ needs to be decreased from 1.40 $M_{\odot}$ until the two curves intersect. Additionally, this implies that $\dot{\gamma}_{tot}$ could also be used to constrain the NS mass in these systems since only a subset of the values of $M_{_{NS}}$ yield a solution for $M_{_{WD}}$ for a given value of $\dot{\gamma}_{tot}$.}

\subsection{System B}

\begin{figure}
\epsscale{1.17}
\plotone{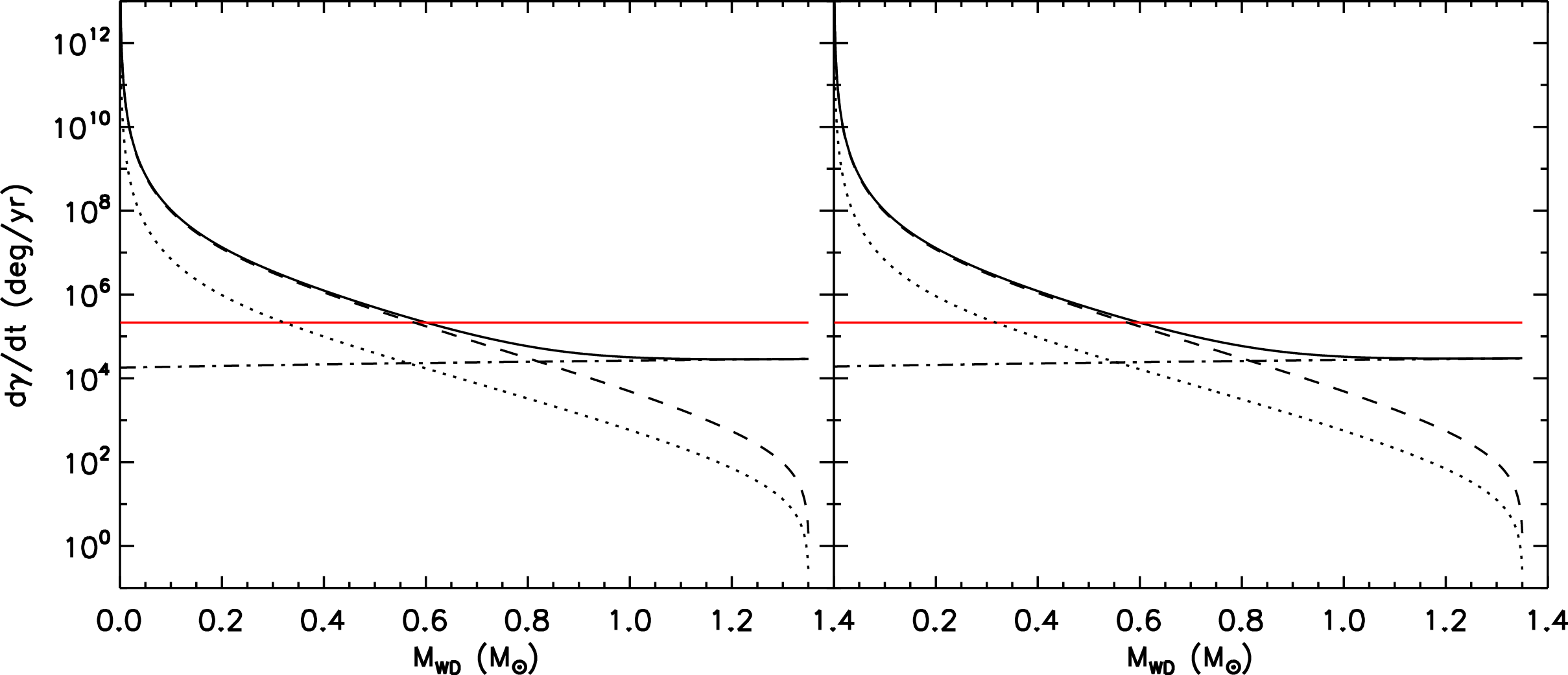}
\caption{System B: $\dot{\gamma}$ as a function of $M_{_{WD}}$ for $M_{_{NS}}$ changed to 1.27 $M_{\odot}$ (left) and 1.40 $M_{\odot}$ (right). The dashed line indicates the tidal contribution, the dotted line indicates the rotational contribution and the dot-dashed line indicates the GR contribution. The solid line represents the total periastron precession rate. The solid red line is the total periastron precession rate we want to match by changing the WD mass.}
 \label{sysB}
\end{figure}

In Fig. \ref{sysB}, we show results for System B similar to those described above for System A. Since this system has a strong tidal contribution, $\dot{\gamma}_{tot}$ decreases as a function of $M_{_{WD}}$ for most of the range in $M_{_{WD}}$ considered. This results in a single solution for $M_{_{WD}}$ for both values of $M_{_{NS}}$. As before, we find that the curve for $\dot{\gamma}_{tot}$ shifts up as $M_{_{NS}}$ is increased, but, in this case, the roots for each NS mass are nearly the same. {A comparison of Figs. \ref{sysA} and \ref{sysB} shows} that as the relative tidal contribution increases the curve for $\dot{\gamma}_{tot}$ spreads out. Thus, even though a second solution for $M_{_{WD}}$ could theoretically exist for systems with $\dot{\gamma}_{Tid}/\dot{\gamma}_{tot} > 0.1$, its value would be greater than the Chandrasekhar limit. 

\clearpage

\end{document}